\documentclass[aps,prb,reprint,showpacs,superscriptaddress,longbibliography]{revtex4-2}
\usepackage{color}
\usepackage[utf8]{inputenc}
\usepackage{graphicx}
\usepackage{xcolor}
\usepackage{braket}
\usepackage{float}
\usepackage{amssymb}
\usepackage{amsmath}
\usepackage[makeroom]{cancel}
\usepackage{upgreek}
\usepackage{commath}
\usepackage{siunitx}
\usepackage[titletoc]{appendix}

\definecolor{Red}{cmyk}{0, 1.0, 0.81, 0.04}
\definecolor{Teal}{cmyk}{0.84,0.0,0.59,0.0}
\usepackage{hyperref}
\hypersetup{colorlinks=true, linkcolor=Red, citecolor=Teal, urlcolor=Teal}

\emergencystretch=\maxdimen
\def\s{\sigma}
\def\ave#1{\langle #1\rangle}
\newcommand{\avs}{\ave{s}}
\newcommand{\iv}{\mathbf{i}}
\newcommand{\jv}{\mathbf{j}}
\newcommand{\qv}{\mathbf{q}}

\newcommand{\di}{i}          %% default math "i"

\begin{document}
\title {Weyl semimetallic, N\'eel, spiral, and vortex states in the Rashba-Hubbard model}

\author{Sebastião dos Anjos Sousa-Júnior}
\affiliation{Department of Physics,
University of Houston, Houston, Texas 77004}
\author{Rubem Mondaini}
\affiliation{Department of Physics,
University of Houston, Houston, Texas 77004}
\affiliation{Texas Center for Superconductivity, University of Houston, Houston, Texas 77204, USA}

\begin{abstract}
We investigate the evolution of magnetic phases in the Hubbard model under strong Rashba spin-orbit coupling on a square lattice. By using Lanczos exact diagonalization and determinant quantum Monte Carlo (DQMC) simulations, we explore the emergence of various magnetic alignments as the ratio between the regular hopping amplitude, $t$, and the Rashba hopping term, $t_R$, is varied over a broad range of Hubbard interaction strengths, $U$. In the limit $t_R \rightarrow 0$, the system exhibits N\'eel antiferromagnetic order, while when $t \sim t_R$, a spiral magnetic phase emerges due to the induced anisotropic Dzyaloshinskii–Moriya interaction. For $t_R > t$, we identify the onset of a spin vortex phase. At the extreme limit $t = 0$($t_R \neq 0 $), we perform finite-size scaling analysis in the Weyl semimetal regime to pinpoint the quantum critical point associated with the spin vortex phase, employing sign-free quantum Monte Carlo simulations -- the extracted critical exponents are consistent with a Gross-Neveu-type quantum phase transition.
\end{abstract}

\maketitle

\section{Introduction}

Spin-orbit coupling is a key phenomenon in condensed matter physics, emerging in systems with broken inversion symmetry. It results in an effective magnetic field experienced by the electrons that arise from spin-momentum locking, driving phase transitions by altering electronic energy bands and generating spin textures in momentum space. Among its various forms, the Rashba spin-orbit coupling (RSOC) is particularly notable, occurring in multiple materials such as heterostructures\,\cite{Nitta1997, Caviglia2010}, surface states of metals\,\cite{LaShell1996} and transition-metal oxides\,\cite{Sunko2017}, and through proximity effects from substrates \cite{Min2006}. While RSOC initially resolves spin-+energy bands in tight-binding models, many other phenomena arising from RSOC have been investigated\,\cite{Manchon2015}. For example, given the tunability of RSOC\,\cite{Schultz1996,King2011}, or by considering spin precession\,\cite{Datta1990, Koo2009}, its applicability to spin devices has been extensively studied. Additionally, the intrinsic spin Hall effect, which is relevant to the field of spintronics, has also been explored in the context of RSOC\,\cite{Murakami2003, Sinova2004}.

While electron-electron interactions do not play a significant role in many of the aforementioned systems, the coexistence of RSOC and strong Coulomb interactions induces exotic magnetic effects via the Dzyaloshinskii-Moriya interaction (DMI)\,\cite{Dzyaloshinsky1958,Moriya1960}. This phenomenon has been predicted\,\cite{Sergienko2006} and observed in materials\,\cite{Seki2012,Nagaosa2013,Wilson2014,Yang2018,Hallal2021}. In addition, emulation of spin-orbit effects in the context of ultra-cold atomic systems was further demonstrated\,\cite{Cheuk2012,Ji2014,Jimenez-Garcia2015} with additional proposals\,\cite{Grusdt2017,Li2018} unraveling its topological properties. The interest in such emulators is owed to their tunability in the interaction strength they offer\,\cite{Gross2017}, which can result in engineered DMI.

From a theoretical perspective, the Hubbard Hamiltonian is one of the simplest models to describe electron-electron interactions, often leading to magnetic transitions and Mott insulating phases\,\cite{Hirsch1985,Paiva2005,Meng2010,Sorella2012}. For example, antiferromagnetism emerges on a half-filled square lattice for any finite Hubbard interaction, $U$, due to a nested Fermi surface and a van Hove singularity\,\cite{Hirsch1985}. In the regime where both Hubbard interaction and RSOC are strong, a spin Hamiltonian can describe the resulting magnetic phases on a square lattice, applicable to both fermionic and bosonic systems\,\cite{Cole2012,Minar2013,Wang2017}. The ground state of the Rashba-Hubbard model (RHM) has been investigated in the weak $U$ regime using random phase approximation and sine-square deformed mean-field theory for intermediate interactions\,\cite{Kawano2023}. Other studies have employed mean-field approximations\,\cite{Kennedy2022,Kubo2024,jain2024} or examined small system sizes\,\cite{Brosco2018,Brosco2020} through exact diagonalization. In most cases, the anisotropic spin interactions induced by RSOC in the RHM on a square lattice result in spiral and vortex phases, which is direct evidence of the  DMI that emerges effectively. 

The effects of RSOC have also been studied within electron-phonon systems, such as in the Holstein model, where it perturbs a charge-density-wave phase instead of an antiferromagnetic state\,\cite{Fontenele2024,Faundez2024}. Furthermore, studies predict that superconductivity emerges in the RHM upon doping, both with repulsive\,\cite{Shigeta2013,Beyer2023} and attractive\,\cite{Tang2014} interactions. On a half-filled square lattice\,\cite{Rosenberg2017,Faundez2024}, quantum Monte Carlo (QMC) simulations demonstrated the onset of coexisting charge order and superconductivity. At the same time, other QMC studies of the repulsive case have discussed the appearance of exotic magnetic alignment on a honeycomb lattice\,\cite{Wan2022}. Despite extensive literature on magnetic ordering in the RHM, results often depend on the employed method. Therefore, an unbiased study of the repulsive case across a wide range of RSOC and Hubbard interaction strengths is necessary to describe its ground state magnetic phase diagram comprehensively. Moreover, accurately determining the transition between the Weyl semimetal, which emerges at large RSOC, and spin vortex phases would shed light on the interplay between topology and strong interactions in the context of spin-orbit coupling.

In this work, we investigate the properties of the repulsive RHM using both the Krylov-Schur-based exact diagonalization method\,\cite{Stewart2002,Slepc} and the determinant quantum Monte Carlo (DQMC) method\,\cite{Blankenbecler1981,Hirsch1985,White1989} on a half-filled square lattice. The layout of the paper is as follows. In Sec.\,\ref{sec:Meth}, we introduce the Hamiltonian and highlight the numerical methods. Section \ref{sec:ED_results} presents the results for the Lanczos method, while the DQMC results are discussed in Sec.\,\ref{sec:DQMC_results}. Finally, Sec.\,\ref{sec:conclusion} summarizes our findings.

\section{Model and Method}
\label{sec:Meth}

The Hamiltonian of the system reads
\begin{align}
    \mathcal{\hat H}=&-t\sum_{\braket{\iv,\jv},\sigma}( \hat c_{\iv,\sigma}^{\dagger}\hat c_{\jv,\sigma}^{\phantom{\dagger}} + \text{H.c.} ) + U\sum_\iv \hat n_{\iv,\uparrow} \hat n_{\iv,\downarrow}  \nonumber \\  
    &-\di t_{R}\sum_{\braket{\iv,\jv},\sigma,
    \sigma^\prime}\hat c_{\iv,\sigma}^{\dagger}(\mathbf{d}_{\iv,\jv}\times \boldsymbol{\hat \sigma})_z^{\sigma,\sigma^\prime}\hat c_{\jv,\sigma^\prime}^{\phantom{\dagger}},
    \label{eq:RHH}
\end{align}
where, $\hat c_{\iv,\s}^{\dagger}$ ($\hat c_{\iv,\s}^{\phantom{\dagger}}$) is the creation (annihilation) operator for an electron with spin $\sigma=\uparrow,\downarrow$ at site $\iv$. The hopping integral between neighbor sites, $\braket{\iv,\jv}$, is given by $t$. The second term describes the screened Coulomb repulsion between electrons on the same site, with $\hat n_{\iv,\s}=\hat c_{\iv,\s}^{\dagger}\hat c_{\iv,\s}^{\phantom{\dagger}}$ being the electron number operator on a site $\iv$ --- we focus on the half-filling case $\sum_{\s}\langle\hat n_{\iv,\s}\rangle=1$. The magnitude of the Rashba spin-orbit coupling is set by $t_{R}$, where  $\mathbf{d}_{ij}$ is the unity vector along the direction connecting site $\iv$ to $\jv$; the $\boldsymbol{\hat \sigma}$ operator is a vector of Pauli matrices $\boldsymbol{\hat \sigma}=(\hat \sigma_{x}, \hat \sigma_{y}, \hat \sigma_{z})$. 

The ground state properties of the RHM are extracted using the Krylov-Schur diagonalization method\,\cite{Stewart2002,Slepc} and determinant quantum Monte Carlo (DQMC) simulations\,\cite{Blankenbecler1981,Hirsch1983,Hirsch1985,rrds2003} -- a benchmark is given in Appendix \ref{App:Bench}. We utilize translational invariance in the former to reduce the original Hilbert dimension of about 600 million states in a $4\times 4$ lattice. In the latter, the non-commuting terms of Eq.\,\eqref{eq:RHH} are separated using the Suzuki-Trotter decomposition, introducing an imaginary time dimension, $L_{\tau} = \beta / \Delta \tau$, where $\beta$ is the inverse temperature, and $\Delta \tau$ establishes a controllable numerical error as $\mathcal{O}(\Delta\tau^2)$. To recast the quartic interaction term in quadratic form, we apply a Hubbard-Stratonovich (HS) transformation\,\cite{Hirsch1983}. This allows one to trace over fermionic degrees of freedom in the partition function, where the remaining degrees of freedom on the auxiliary fields are then evaluated via importance sampling, with a determinant serving as the configuration weight. For finite $t_R$, the spin-flip term in \eqref{eq:RHH} prevents the configuration weight from being expressed as a product of up and down determinants, as often accomplished for Hamiltonians with spin-decoupled hoppings.

As such, unlike what occurs for $t_R=0$, where a particle-hole transformation can be used to prove the absence of the sign problem at half-filling\,\cite{Hirsch1985,White1988,Loh1990,Mondaini2022,Mondaini2023}, at finite $t_R$, simulation results suffer from increasing statistical noise, as the single determinant appearing in the partition function is no longer positive definite. This leads to an exponential decrease in their average sign, $\avs$, with both the system size and the inverse temperature $\beta$~\cite{Loh1990}. The behavior as a function of $\theta$ and $U$ is presented in Appendix~\ref{App:sign_prob}; it restricts accurate simulations to small $U$ values, where the sign problem is manageable. In the extreme limit $t\to 0$, an SU(2) gauge transformation can be implemented on the Rashba-hopping term such that it is transformed into a spin-independent hopping term in the presence of a $\pi$-flux per plaquette\,\cite{Kawano2023, Wan2022}, that does not manifest a sign problem\,\cite{Parisen2015,Otsuka2016}. As a result, it permits a finite-size scaling analysis of pertinent physical quantities, particularly those that quantify magnetism in this regime.

\begin{figure*}[t]
    \centering
    \includegraphics[scale=0.5]{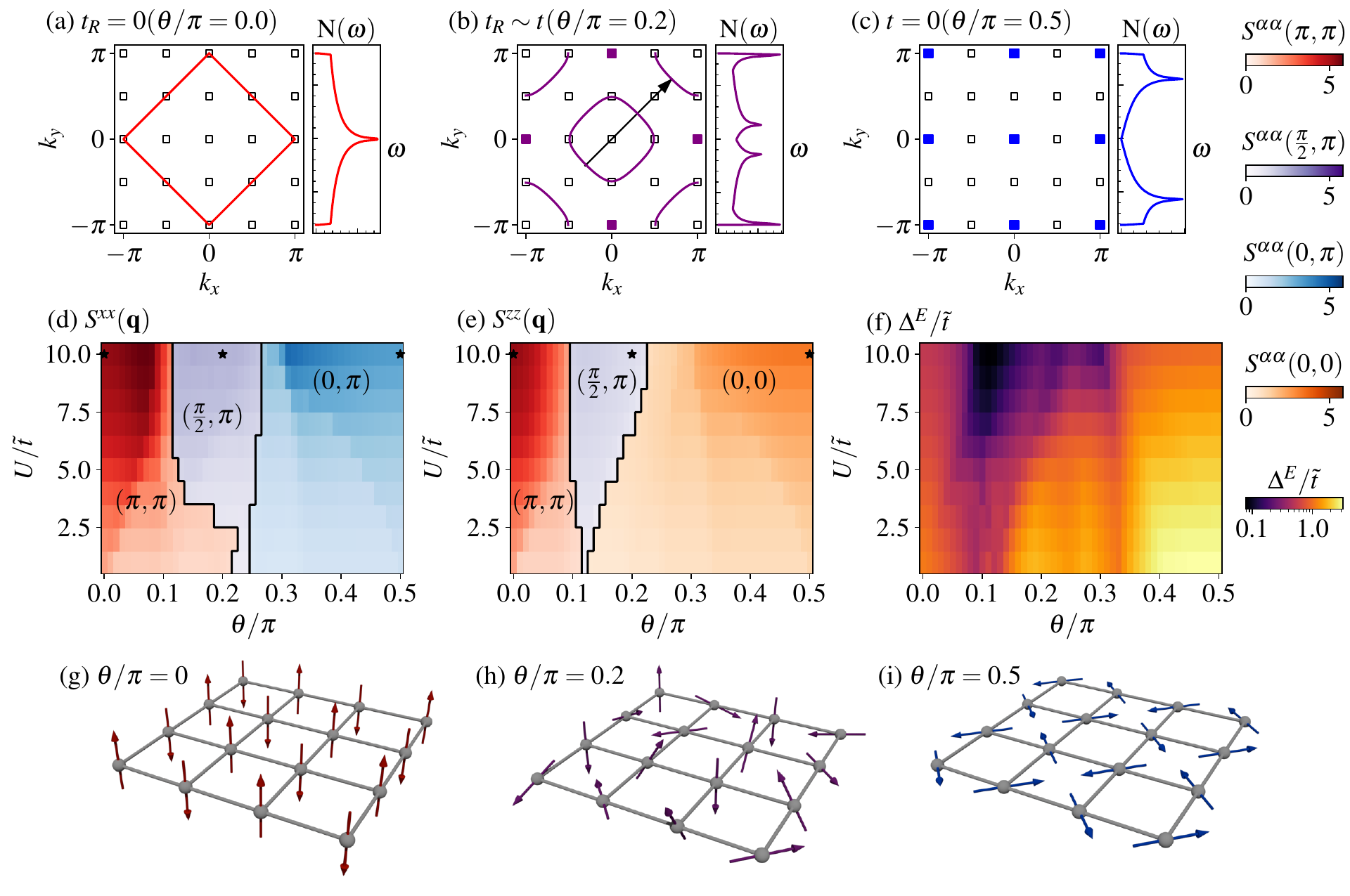}
    \caption{Fermi surface and density of states for $\theta/\pi = 0.0$ (a), $\theta/\pi = 0.2$ (b), and $\theta/\pi = 0.5$ (c), using the parameterization $t = \tilde{t} \cos{\theta}$ and $t_R = \sqrt{2} \tilde{t} \sin{\theta}$. Contour plots of the spin structure factor for the $x$ (d) and $z$ (e) components are presented in the $\theta\times U$ space for an $L=4$ lattice after averaged twisted boundary conditions. The different color schemes map different magnetic wave vectors ${\bf q}$, while their intensity indicates the corresponding structure factor, $S^\alpha({\bf q})$ -- see color bars. (f) The spectral energy gap, $\Delta^E / \tilde{t}$, between the ground and the first excited states. Panels (g), (h), and (i) depict schematic representations of the magnetic order corresponding to the black stars in panels (d) and (e), identifying the (g) Néel, (h) spiral, and (i) vortex phases. For the spiral phase (h), the complex magnetic alignment is derived from spin correlations using the relation $\mathbf{s}_\iv \propto \langle S^z_0 {\bf S}_\iv \rangle$\,\cite{Wan2022}.
    }
    \label{fig:ED_diagram}
\end{figure*}

In both Krylov-Schur and DQMC methods, we investigate the magnetic properties of the system through the spin-spin correlation functions:
\begin{equation}
	S^{\alpha\beta}(\textbf{i},\textbf{j} )=   \langle  \hat S^\alpha_{\textbf{i}}  \hat S^\beta_{\textbf{j}}  \rangle, 
\end{equation}
where $\{ \alpha,\beta\} = {x,y,z}$. Due to the anisotropy of the emergent DMI in the resulting spin Hamiltonian in the strongly interacting limit\,\cite{MacDonald1988,Cole2012,Minar2013,Wang2017}, we probe the occurrence of different magnetic alignments by analyzing the behavior of the spin structure factor for each component:
\begin{equation}
S^{\alpha\alpha}(\textbf{q}) = \frac{1}{N} \sum_{\textbf{i}, \textbf{j}} e^{-i\textbf{q} \cdot \mathbf{d}_{\iv,\jv}} S^{\alpha\alpha}(\textbf{i},\textbf{j} ),
\label{eq:spinsf}
\end{equation}
where distinct magnetic wave vectors, $\qv$, signal the emergence of N\'eel, spiral, or vortex phases, and $N = L \times L$ is the number of sites.

To estimate the critical regions, we use the spin correlation ratio of the average spin structure factor, $S(\textbf{q}) = (S^{xx}(\textbf{q})+S^{yy}(\textbf{q})+S^{zz}(\textbf{q}))/3$:
\begin{equation}
    R(L) = 1 - \dfrac{ S(\textbf{q} +\delta \textbf{q}) }{S(\textbf{q})},
    \label{eq:corratio}
\end{equation}
where  $ | \delta \textbf{q}| = 2\pi/L$ is the smallest momentum displacement allowed at a given lattice size.  For sufficiently large $L$, the correlation ratio $R(L,U)$ follows a finite-size scaling form\,\cite{Fisher71,dosSantos81a,Barber83}:
\begin{equation}
 R(L,U)=\mathcal{F}[(U-U_c)\,L^{1/\nu}],
  \label{eq:RFSS}
\end{equation}
where $\mathcal{F}$ is a universal function, and $\nu$ is the correlation length critical exponent. At the critical point, $\mathcal{F}(0)$ is a constant independent of $L$, ensuring that curves of $R$ for different lattice sizes cross at the critical point. This crossing provides an increasingly precise estimate of the critical value as $L\to\infty$. Therefore, finite-size scaling analyses of the crossing of $R(L)$ for different lattice sizes provide estimates for the location of the critical region\,\cite{Kaul2015} since this quantity is renormalization-group invariant.

\begin{figure}[t]
    \centering
    \includegraphics[scale=0.5]{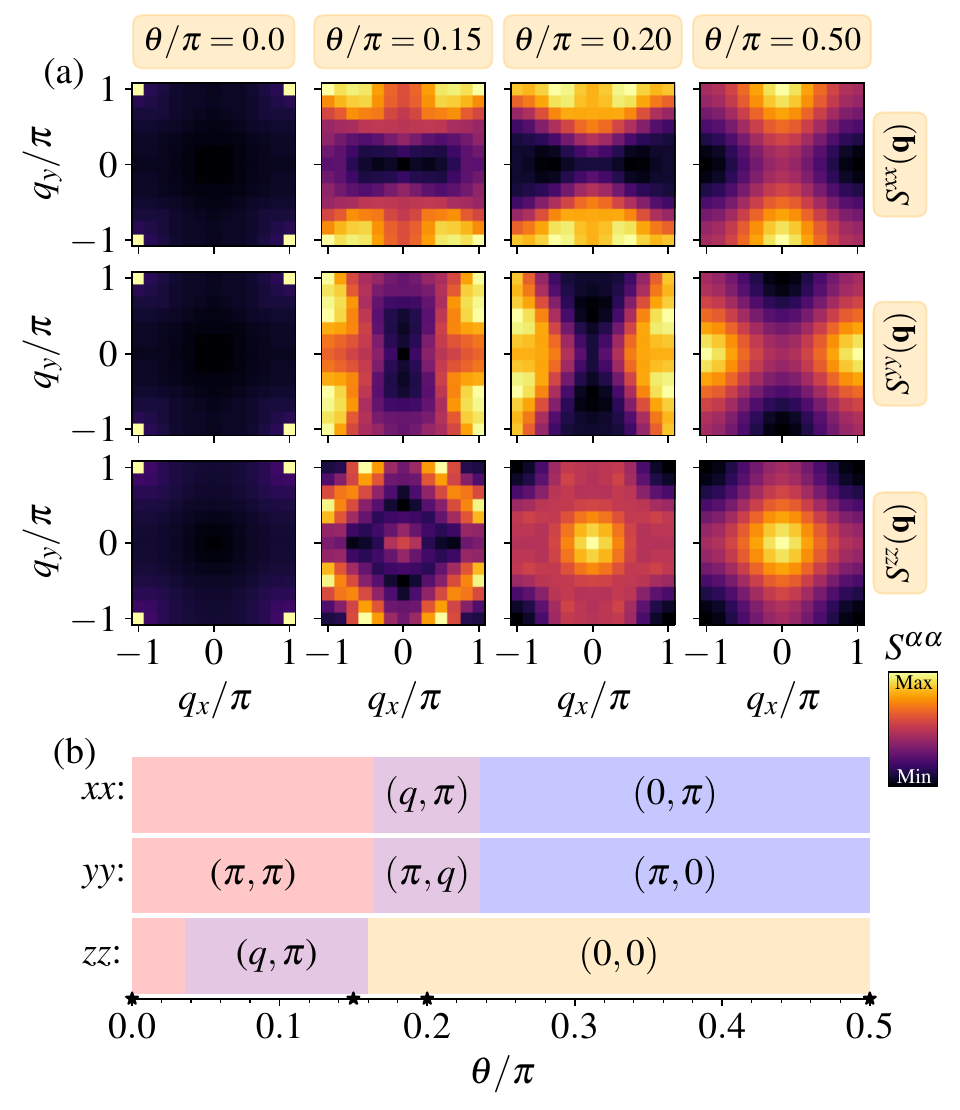}
    \caption{(a) Contour plots of the spin structure factors for each spin component, calculated for $L = 12$, $\beta \tilde{t} = 24$, $U/\tilde{t} = 2$, and varying values of $\theta$. (b) Wave vector corresponding to the maximum structure factor for each spin component as a function of $\theta/\pi$; the star markers depict the values of $\theta$ used in (a). Note that the wave vector ${\bf q}$ is always commensurate, with $q$ in the spiral phases with peaks at $(q,\pi)$ or $(\pi, q)$ spanning the range $[4\pi/L, 10\pi/L]$. 
    }
    \label{fig:spiral_vortex}
\end{figure}

\section{Exact Diagonalization Results}
\label{sec:ED_results}

We begin our discussion by revisiting the effects of spin-orbit coupling on the spectral properties in the non-interacting case. To this end, we parameterized the hopping integral as  $t = \tilde{t} \cos{\theta} $ and the Rashba hopping as $t_R = \sqrt{2} \tilde{t} \sin{\theta} $, such that the non-interacting bandwidth, $W = 8\tilde{t}$, remains constant for any $\theta$. In so doing, it allows one to easily compare different magnetic regimes by varying $\theta$. In the absence of Rashba hopping $(t_R =0$ for $ \theta = 0)$, the Fermi surface is perfectly nested, and the density of states exhibits a van Hove singularity at half-filling, as displayed in Fig.\,\ref{fig:ED_diagram}(a). These features drive the system to a N\'eel phase for any finite value of $U/\tilde{t}$\,\cite{Hirsch1985}. 

When increasing $\theta$, and thus the $t_R/t$ ratio, the van Hove singularity moves away from half-filling while maintaining the nested Fermi surface and metallic behavior for $U = 0$, as illustrated in Fig.\,\ref{fig:ED_diagram}(b) for $t_R \sim t$ ($\theta/\pi=0.2$). For the pure Rashba case, where $t =0$ ($\theta/\pi = 0.5$), nesting is lost, and the system undergoes a transition from metal to a Weyl semimetal, with a vanishing density of states at the Fermi level, as shown in Fig.\,\ref{fig:ED_diagram}(c). In this latter case, the Fermi level lies on the time-reversal-symmetry protected Weyl points ${\bf k} = [(0,0),(0,\pm \pi ),(\pm \pi, 0),(\pm \pi, \pm \pi)]$. In the vicinity of these points, the energy bands acquire a linear dispersion relation, like those in the Dirac cones in graphene.  

As displayed in Figs.\,\ref{fig:ED_diagram}(a), \ref{fig:ED_diagram}(b), and \ref{fig:ED_diagram}(c), there are only three values of $\theta$ where the Fermi surface for the thermodynamic limit (colored lines) coincides with the allowed ${\bf k}$ points of the $4 \times 4$ square lattice (empty black squares). Therefore, we use twisted averaged boundary conditions\,\cite{Niu1985,Poilblanc1991,Gros1992} to mitigate the finite size effects of the kinetic terms of the Hamiltonian and obtain the spin structure factors shown in Fig.\,\ref{fig:ED_diagram}. In tandem with the drastic change in the Fermi surface and density of states of the system as we tune $\theta$, the magnetic alignment is also altered by the RSOC. 

In addition to the $(\pi,\pi)$ N\'eel phase, the RHM exhibits other spin arrangements, as displayed in Fig.\,\ref{fig:ED_diagram}(d) and \ref{fig:ED_diagram}(e). For $\theta = 0.2\pi$ ($t \sim t_R$), the system displays a spiral phase, as the maximum value of $S^{xx}({\bf q})$, $S^{yy}({\bf q})$, and $S^{zz}({\bf q})$ corresponds to ${\bf q} = (\pi/2,\pi)$ [or its degenerate counterpart, $(\pi,\pi/2)$]. At this point, we highlight the anisotropy of the spin correlations arising from the DMI, as the maximum values of $S^{\alpha\alpha}({\bf q})$ occur not only for different wave vectors $\qv$, but also with different intensities [see, e.g., Fig.\,\ref{fig:spiral_vortex}]. Although nesting is preserved for $\qv = (\pi,\pi)$, the magnetic alignment also changes due to additional contributions from different processes at finite $\theta$, where other vectors contribute to connecting points at the Fermi level, as discussed in Ref.\,\cite{Kawano2023}.

For $0.3 \lesssim \theta \leq 0.5$, a spin vortex phase emerges, characterized by maximum values for $S^{xx}(0,\pi)$ and $S^{zz}(0,0)$, as displayed in Fig.\,\ref{fig:ED_diagram}(d) and (e), respectively. While the distinct magnetic alignments schematically represented in Fig.\,\ref{fig:ED_diagram}(g), (h), and (i) appear over reasonably wide ranges of $\theta$, the N\'eel $(\pi,\pi)$ phase is notably favored for small $U/\tilde{t}$ due to nesting. As the Hubbard interaction strengthens, the transition lines between different magnetic phases evolve into vertical lines, becoming insensitive to $U/\tilde{t}$ values. In the $U \gg \tilde{t}$ limit, the system can be effectively described by a spin Hamiltonian, and our results align with those in Refs.\,\cite{Cole2012,Wang2017} [see Appendix\,\ref{App:IUL}]. Across the $U \times \theta$ parameter space, the spectral energy gap, $\Delta^E/\tilde{t}$, between the ground and the first excited states, also serves as an indicator of the different magnetic phases, with smaller values within the spiral phase (an indication of the competing states in this regime), while the vortex and $(\pi,\pi)$ phases yield larger many-body gaps -- see Fig.\,\ref{fig:ED_diagram}(f).

As a final remark on the ED results, we note that comparing them with the equivalent spin Hamiltonian in the $U \gg \tilde{t}$ limit involves investigating the skyrmion phase, which has been reported to exhibit a $3\times 3$ periodicity\,\cite{Cole2012,Wang2017}. However, such an investigation lies beyond the applicability of our ED calculations on a $4\times 4$ lattice [see discussion in Appendix\,\ref{App:IUL}].

\section{Quantum Monte Carlo results}
\label{sec:DQMC_results}

We now discuss the DQMC results by analyzing the finite-size effects on the different magnetic phases. This approach encounters increasing numerical noise due to the sign problem\,\cite{Hirsch1985,White1988,Loh1990} [see Appendix\,\ref{App:sign_prob}]. However, DQMC simulations for small $U$ are not significantly affected by the sign problem and provide reliable results for the correlation functions. Figure \ref{fig:spiral_vortex} shows the spin correlation functions for fixed $U/\tilde{t}=2$, $\beta\tilde{t}=24$, and $L=12$. 

Similar to what is observed in the ED results, Fig.\,\ref{fig:spiral_vortex}(a) clearly illustrates that the magnetic correlations become highly anisotropic, with peaks in the spin structure factors corresponding to different wave vectors for each spin component once $\theta$ (and thus the RSOC) is varied; the larger lattice size permits, however, a finer momentum resolution. A compilation of the leading wave vectors for each channel is shown in Fig.\,\ref{fig:spiral_vortex}(b), further highlighting the evolution of the magnetic wave vector as we tune $\theta$. In particular, the spiral phases (considering all components) are broader in the $\theta$ range than those observed in the ED results for $L=4$: the onset of the $(q,\pi)$ phase occurs at $\theta/\pi \sim 0.16$ for the $x$ component and at $\theta/\pi \sim 0.04$ for the $z$ component. Despite these small quantitative differences, the $12 \times 12$ lattice results are qualitatively similar to those obtained for the $4 \times 4$ lattice; the differences can be attributed to the fact that larger lattices allow more commensurate wave vectors, effectively enlarging the $(q,\pi)$ phases.

\begin{figure}[t]
    \centering
    \includegraphics[scale=0.5]{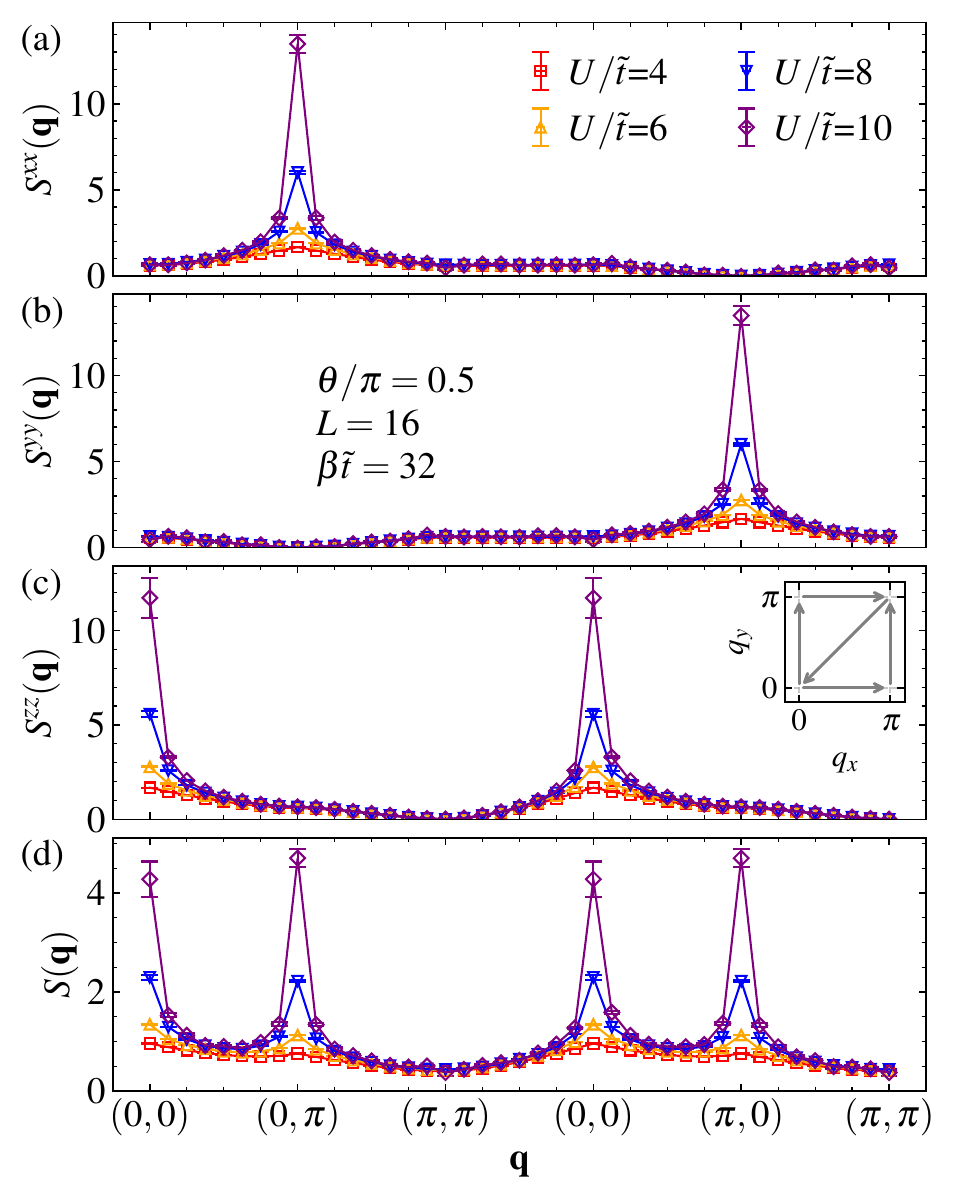}
    \caption{Spin structure factor as a function of the wave vector for different values of $U$, with fixed parameters $L = 16$ and $\beta\tilde{t} = 32$. The panels illustrate the individual contributions from the (a) $x$, (b) $y$, and (c) $z$ components. The inset shows the selected path within the first Brillouin zone. (d) Average spin structure factor, computed as described in Eq.\,\eqref{eq:spinsf}.
    }
    \label{fig:Sq_l16_b32}
\end{figure}

While earlier static\,\cite{Kennedy2022,Kawano2023} and dynamical\,\cite{Zhang2015} mean-field calculations predicted a finite critical interaction $U_c$ for the onset of magnetism with $t_R>0$, a recent study argued that the nested Fermi surface should stabilize antiferromagnetic order in the ground state for any finite $U_c$\,\cite{Kubo2024}. The order parameter, however, is predicted to be exponentially small in this latter scenario. Such weak magnetism could manifest as increasing values of the inverse temperature $\beta$ for observing the saturation of spin-structure factors within our DQMC approach to reach a temperature representative of the system's ground state. However, given the qualitative agreement between the ED and DQMC results, temperature scales do not appear to pose a limitation in determining the dominant magnetic alignment across the entire $U \times \theta$ parameter space. Yet, the emergence of the sign problem prevents a precise scaling analysis for $\theta \in (0,\pi/2)$ to determine the value of $U_c(\theta)$, but qualitatively supports that antiferromagnetic occurs with $U_c\to 0^+$ --- see Appendix\,\ref{App:sign_prob}. Notably, only in the pure Rashba hopping case ($\theta=\pi/2$) does nesting disappear, turning the system to a Weyl semimetal phase for small $U/\tilde{t}$.

\begin{figure}[t]
    \centering
    \includegraphics[scale = 0.5]{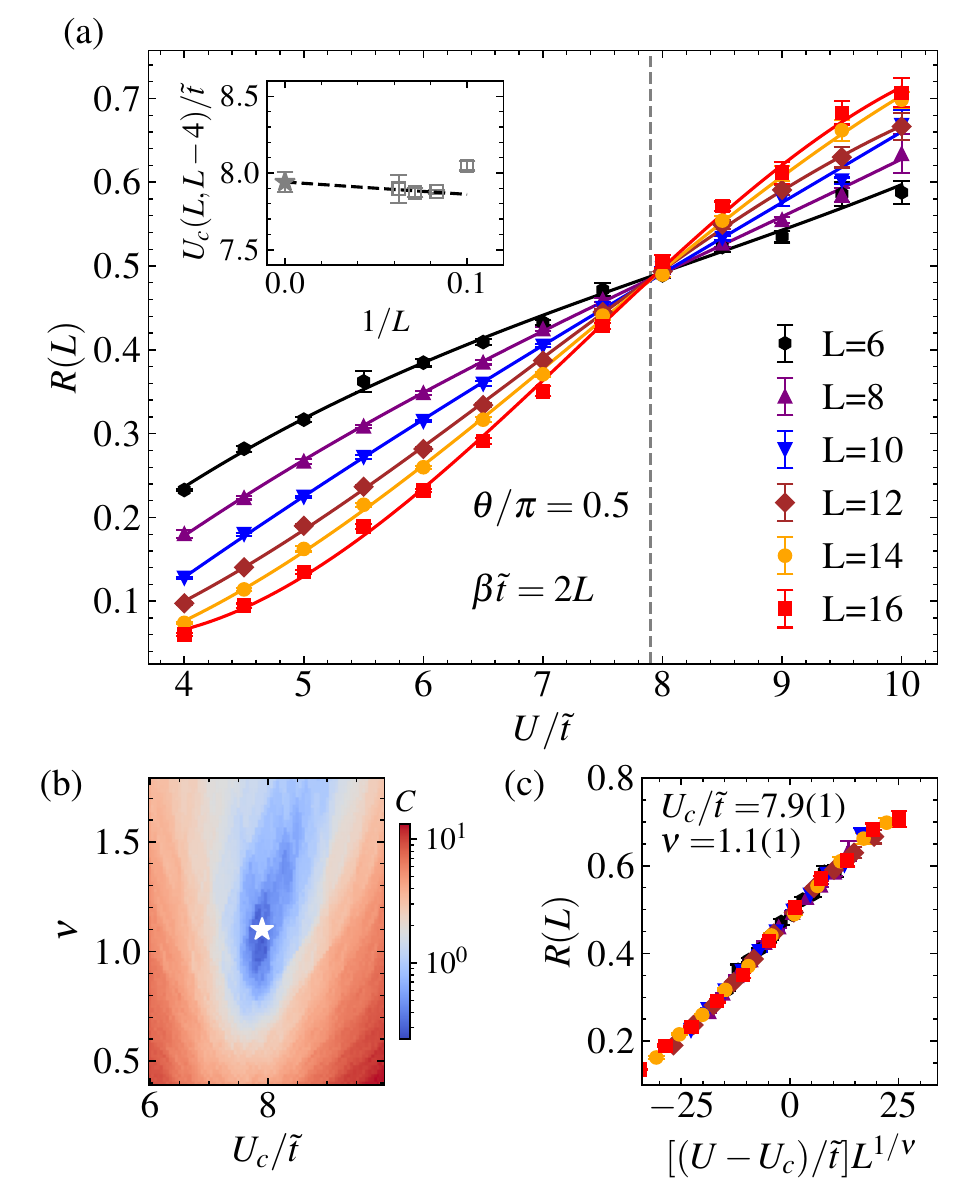}
    \caption{(a) Correlation ratio as a function of $U$ for various lattice sizes, with fixed parameters $\beta = 2L$ and $\theta = \pi/2$. The inset displays the extrapolation of the critical value of $U$ in the thermodynamic limit. (b) Contour plot of the cost function as a function of $\theta$ and $U$. (c) Data collapse of the correlation ratio, highlighting the universal behavior.
    }
    \label{fig:Rc_pure_rashba}
\end{figure}

Focusing on this case, we now examine the transition from the Weyl semimetal to the spin vortex phase at $\theta=\pi/2$. In this pure Rashba limit, the spin structure factors for each component exhibit a continuous increase with $U$, signaling a second-order phase transition. Figure~\ref{fig:Sq_l16_b32} presents the evolution of the structure factors for each spin component as the Hubbard interaction increases. For larger values of $U/\tilde{t}$, the spin structure factors for the $x$, $y$, and $z$ components develop pronounced peaks at specific wave vectors, $\qv$, as shown in Fig.\,\ref{fig:Rc_pure_rashba}(a), (b), and (c), respectively. While the $z$ component displays a notable peak at $\qv = (0,0)$, the correlation ratio for this wave vector (not shown) does not effectively indicate the transition to the coplanar spin vortex phase. Instead, Fig.\,\ref{fig:Sq_l16_b32}(d) shows that the average spin structure factor reaches its maximum value at $\qv = (0,\pi)$ and $\qv = (\pi,0)$, which defines the wave vector used to compute the correlation ratio as in Eq.\,\eqref{eq:corratio}. 

To estimate the critical value of $U/\tilde{t}$ for the Weyl semimetal to spin vortex transition, we compute $R(L)$ for $\qv = (0,\pi)$. The results for the correlation ratio as a function of $U/\tilde{t}$ are presented in Fig.\,\ref{fig:Rc_pure_rashba}(a). As the system size increases, the values of $R(L)$ decrease (increase) with the lattice size for interaction values greater (lower) than $U/\tilde{t} \sim 8$. To determine the position of the crossing points as the system size grows, we perform a finite-size scaling analysis by extrapolating the crossing points between curves for lattices of sizes $L$ and $L-4$. These define the critical values of the Hubbard interaction, $U_c(L,L-4)$, as shown in the inset of Fig.\,\ref{fig:Rc_pure_rashba}(a). By taking the $1/L \to 0$ limit, a linear fit yields the critical point: $U/\tilde{t} = 7.94 \pm 0.07$.

To precisely determine the critical point, we independently perform a data collapse of the correlation ratio by minimizing the cost function $C(\nu, U_c) = \sum_j (|y_{j+1} - y_j|)/(\max\{y_j\} - \min\{y_j\})-1$\,\cite{Suntajs2020}, where $y_j$ are the values of $R(L)$ ordered according to their corresponding $[U-U_c)/\tilde t]L^{1/\nu}$, quantifying the distance between consecutive points in the collapse: its minimum identifies the best collapse. The results for the cost function are shown in Fig.\,\ref{fig:Rc_pure_rashba}(b), where a white star marks the parameters that minimize the distances between consecutive points. The corresponding collapsed correlation ratios are displayed in Fig.\,\ref{fig:Rc_pure_rashba}(c), yielding critical values of $U_c/\tilde{t} = 7.9 \pm 0.1$ and $\nu = 1.1 \pm 0.1$.  Therefore, the critical points derived from the extrapolation of the crossing points of $R(L)$ and the minimization of $C$ are in agreement. The dashed line in Fig.\,\ref{fig:Rc_pure_rashba}(a) highlights this consistency, emphasizing the concordance between the collapse and crossing methods. Finally, the critical exponent $\nu$ obtained here is consistent with results from other models in the Gross-Neveu universality class\,\cite{Assaad2013,Otsuka2016,Parisen2015,Tang2018}, as one would expect from the gauge transformation that maps this model at $\theta = \pi/2$ to the Hubbard model with a spin-independent hopping term in the presence of a $\pi$-flux per plaquette\,\cite{Kawano2023, Wan2022}.

A direct evidence of the onset of the spin-vortex phase is the emergence of a gap for single-particle excitations, as shown in Appendix\,\ref{App:spectral}. Additionally, the topological characteristics of the Weyl state and its suppression with increased Hubbard $U$, can also be inferred via the Berry curvature computed at small cluster sizes (see Appendix\,\ref{App:berry_curv}), indicating the spin vortex phases to be topologically trivial.
\section{concluding remarks }
\label{sec:conclusion}

In summary, we investigate the evolution of magnetic phases in the Rashba-Hubbard model on a square lattice using Krylov-Schur exact diagonalization (ED) and determinant quantum Monte Carlo. By tuning the ratio between the strengths of regular hopping and spin-flip Rashba hopping, the emerging anisotropic Dzyaloshinskii–Moriya interaction disrupts the N\'eel phase, driving the system into spiral and spin vortex phases. For weak interactions, nesting enhances the antiferromagnetic $(\pi,\pi)$ phase, whereas stronger screened Coulomb repulsion stabilizes the spiral and spin vortex phases over wider ranges of Rashba coupling in the ED calculations. For $U/\tilde{t}=2$, the DQMC approach remains free from severe sign problems and produces results consistent with the ED calculations. At strong interaction limits, our ED results qualitatively align with those in Refs.\,\cite{Cole2012,Wang2017}, although we are unable to access the skyrmion phase owing to the small system size amenable to this type of calculation.

In the extreme case where only Rashba hopping is present, the system becomes a Weyl semimetal and transitions to a spin vortex phase with growing $U$ [see cartoon in Fig.\,\ref{fig:ED_diagram}(i)], wherein a critical exponent is extracted consistent with the Gross-Neveu universality class. Our findings can potentially provide insights into the interplay between strong spin-orbit coupling and screened (local) Coulomb interactions, contributing to understanding magnetic behavior in a broad class of materials where both are relevant.

\section*{ACKNOWLEDGMENTS}
S.A.S.-J. gratefully acknowledges financial support from the Brazilian Agency CNPq. R.M.~acknowledges support from the T$_{\rm c}$SUH Welch Professorship Award. Part of the calculations used resources from the Research Computing Data Core at the University of Houston. This work also used TAMU ACES at Texas A\&M HPRC through allocation PHY240046 from the Advanced Cyberinfrastructure Coordination Ecosystem: Services \& Support (ACCESS) program, which is supported by U.S. National Science Foundation grants 2138259, 2138286, 2138307, 2137603, and 2138296.

%%%%%%%%%%%%%%%%%%%%%%%%%%%%%%%%%%%%%%%%%%%%%%%%%%%%%%%%%%%%%%%%%%%%%%%%%%%%%%%%%%%%%
%%%%%%%%%%%%%%%%%%%%%%%%%%%%%%%%    Appendix     %%%%%%%%%%%%%%%%%%%%%%%%%%%%%%%%%%%%
%%%%%%%%%%%%%%%%%%%%%%%%%%%%%%%%%%%%%%%%%%%%%%%%%%%%%%%%%%%%%%%%%%%%%%%%%%%%%%%%%%%%%
\appendix

%%%%%%%%%%%%%%%%%%%%%%%%%%%%%%%%%%%%%%%%%%%%%%%%%%%%%%%%%%%%%%%%%%%%%%%%%%%%%%%%%%%%%
\section{Data comparison for the pure Rashba limit}
\label{App:Bench}

\begin{figure}[t]
    \centering
    \includegraphics[scale=0.5]{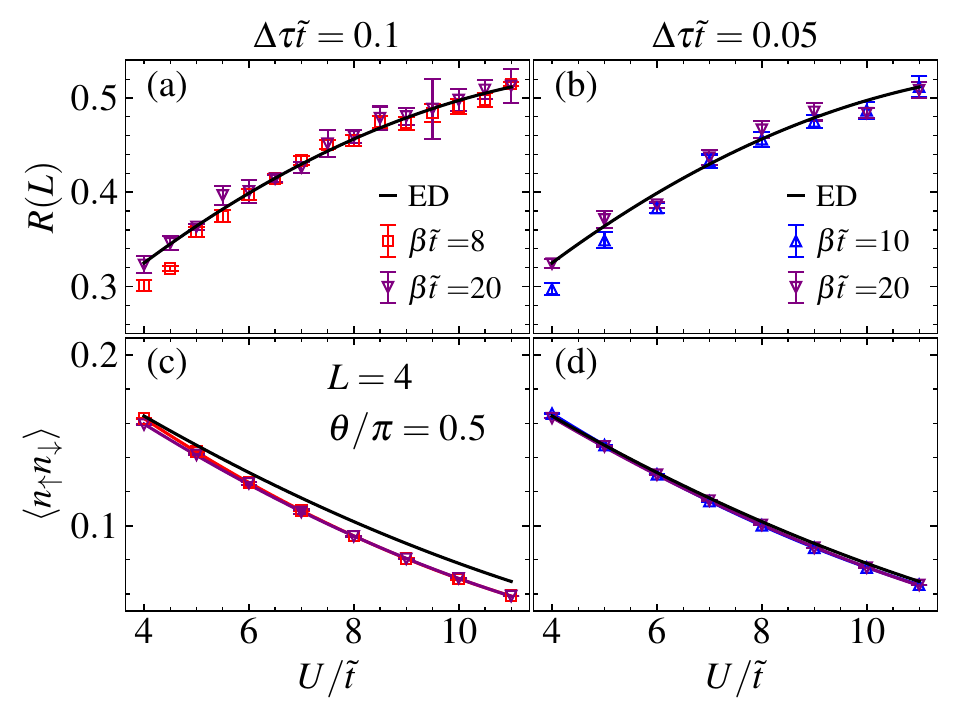}
    \caption{Correlation ratio (upper panels) and double occupancy (lower panels) as functions of $U/\tilde{t}$. The colored empty markers represent DQMC data at different temperatures, while the black solid lines display results from exact diagonalization. The left panels show DQMC data for $\Delta\tau = 0.1$, and the right panels present results for a fixed $\Delta\tau = 0.05$. All data correspond to a $4 \times 4$ lattice with periodic boundary conditions and $\theta = \pi/2$.
    }
    \label{fig:ed_comp}
\end{figure}

As a benchmark for the simulations, we first perform an accuracy test by comparing data extracted from the ED and DQMC methods. It serves a dual purpose: (i) test the influence of the single controllable error in the DQMC calculations, the finite imaginary-time (Trotter) discretization $\Delta \tau$, and (ii) examines the effects of temperature on calculating the physical quantities analyzed in this study. Figure \ref{fig:ed_comp} summarizes the results for the pure Rashba limit ($\theta/\pi = 0.5$). Panels (a) and (b) in Fig.\,\ref{fig:ed_comp} display the correlation ratio as a function of $U/\tilde{t}$ for $\Delta\tau\tilde{t} = 0.1$ and $\Delta\tau\tilde{t} = 0.05$, respectively. Apart from slight discrepancies at smaller values of $U$, the curves for DQMC data are consistent across all temperatures and align well with the zero-temperature results from ED for the correlation ratio at ${\bf q} = (0,\pi)$. This consistency validates the assumption of zero-temperature projection in the analysis of the correlation ratio presented in Sec.\,\ref{sec:DQMC_results}.

For completeness, we also examine the dependence of the Trotter discretization on the double occupancy, $\langle \hat n_\uparrow \hat n_\downarrow \rangle$, which decreases as local moment formation becomes more prominent. Figures \ref{fig:ed_comp}(c) and (d) present the results for $\langle \hat n_\uparrow \hat n_\downarrow \rangle$ as a function of $U/\tilde{t}$ for two values of $\Delta \tau$. Unlike the correlation ratio data, the double occupancy more strongly depends on $\Delta\tau\tilde{t}$ values. Notably, only the data shown $\Delta \tau \tilde t = 0.05$ exhibit full compatibility between ED and DQMC results. 

%%%%%%%%%%%%%%%%%%%%%%%%%%%%%%%%%%%%%%%%%%%%%%%%%%%%%%%%%%%%%%%%%%%%%%%%%%%%%%%%%%%%%
\section{Sign Problem}
\label{App:sign_prob}

\begin{figure}[t]
    \centering
    \includegraphics[scale=0.5]{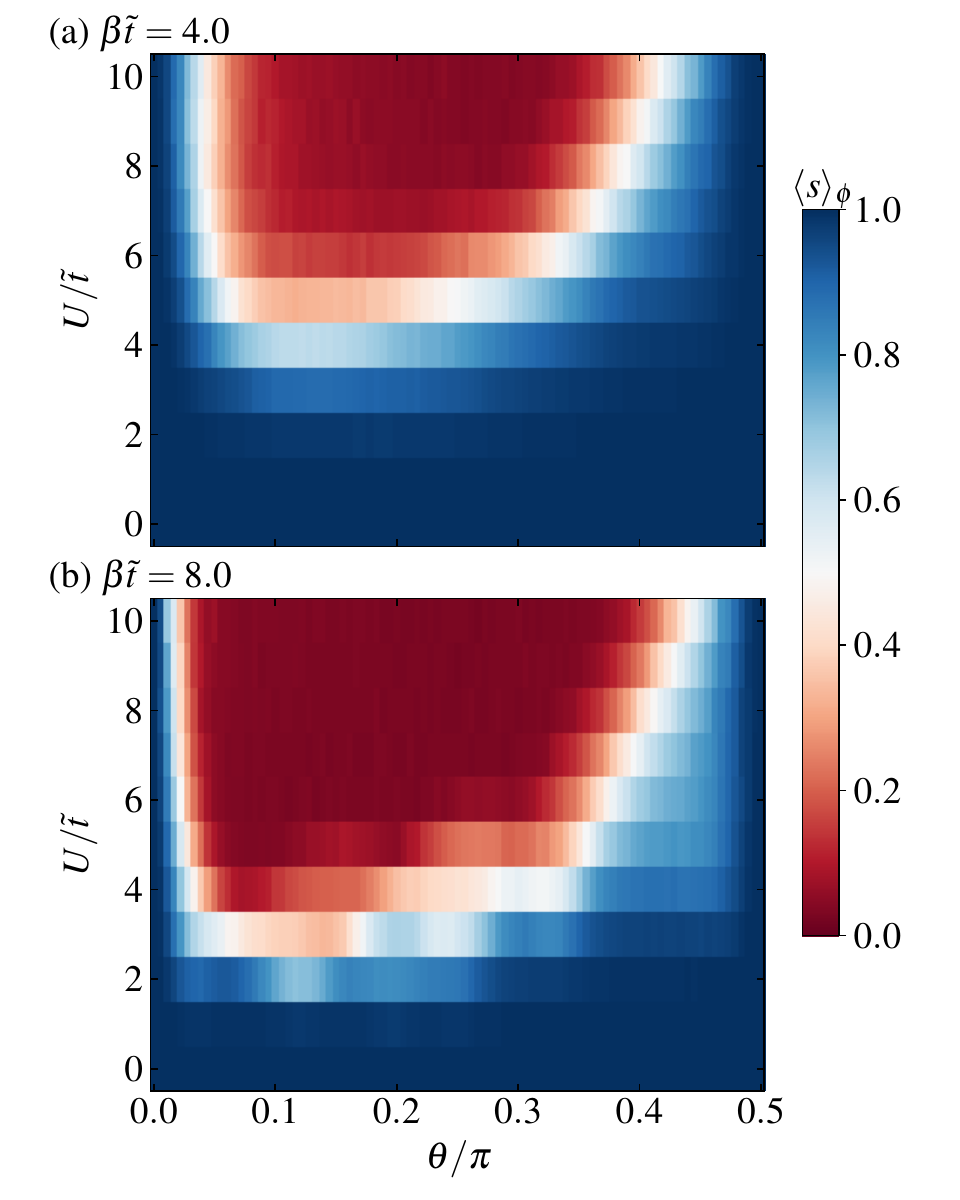}
    \caption{$U \times \theta$ contour plot of the average determinant sign for $L=4$, with (a) $\beta \tilde{t} = 4$ and (b) $\beta \tilde{t} = 8$. Each point is averaged over sixteen twisted boundary conditions.
    }
    \label{fig:sign_map}
\end{figure}

Including Rashba spin-orbit coupling (RSOC) in the Hubbard Hamiltonian renders the matrices used in the DQMC method complex. The absence of particle-hole symmetry introduces a phase problem in the DQMC simulations\,\cite{Mondaini2023}, leading to a vanishing signal-to-noise ratio in the average estimations as $\beta, L \rightarrow \infty$. We analyze the behavior of the average determinant sign, $\langle s \rangle$, as a function of $\theta$ and $U$. Aside from the sign-problem-free cases ($t_R = 0$ for $\theta = 0$ and $t = 0$ for $\theta = \pi/2$), where $\langle s \rangle = 1$, the average determinant sign monotonically decreases at low temperatures, as shown in Fig.\,\ref{fig:sign_map}. Yet, the slower decrease observed for small $U$ values allows simulations for $L=12$, $\beta\tilde{t}=24$, and $U/\tilde{t}=2$ used in the main text -- see Fig.~\ref{fig:spiral_vortex}. In this context, the symbol $\phi$ denotes averages performed over twisted boundary conditions, which help reduce the finite-size effects and mitigate the phase problem to some extent.

\begin{figure}[t]
    \centering
    \includegraphics[scale=0.5]{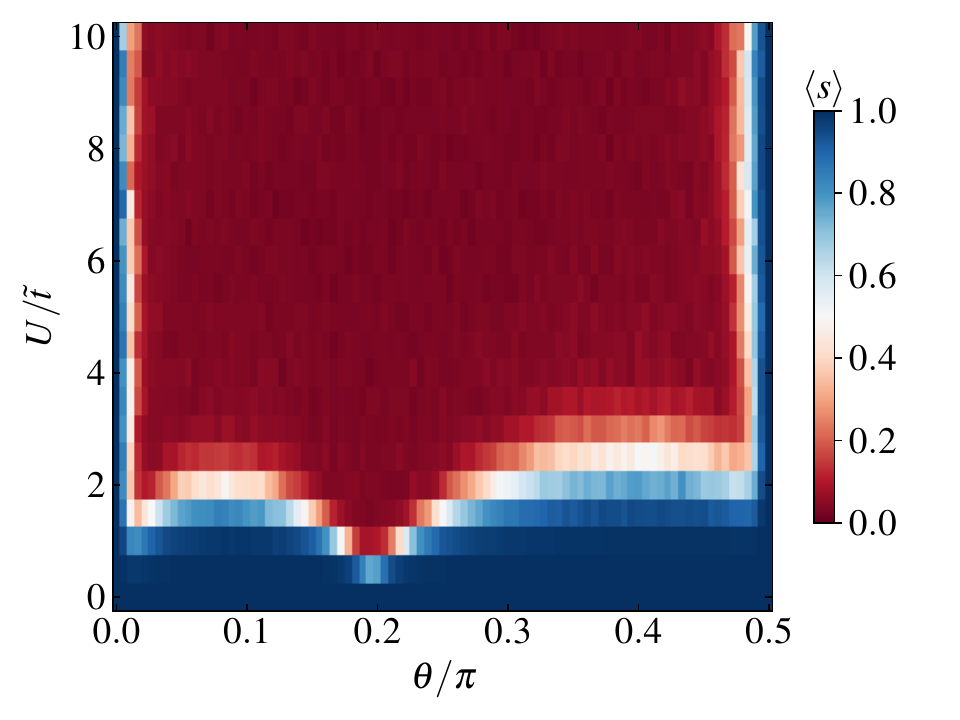}
    \caption{$U \times \theta$ contour plot of the average determinant sign for $L=4$, with $\beta \tilde{t} = 20$ and periodic boundary conditions.
    }
    \label{fig:sign_nesting}
\end{figure}

An indication of these finite-size effects can be clearly seen in Fig.~\ref{fig:sign_nesting} where we report similar data \textit{without} boundary condition averaging, i.e., in the regime of periodic boundary conditions ($\phi=0$) for an $L=4$ lattice. Apart from the symmetry-protected limits $\theta=0$ or $\pi/2$, the sign problem is particularly deleterious at $\theta/\pi \simeq 0.2$, even at small interaction strengths. Non-coincidentally, this is the value at which, for this lattice size, $(\pi,\pi)$-nesting is commensurate with the available momentum points -- see Fig.\ref{fig:ED_diagram}(b). This behavior resembles another model that exhibits nesting at half-filling when connecting two bands, the bilayer Hubbard model\,\cite{Mou2022}. There, although no sign problem occurs because the configuration weights are split for each spin component, and by symmetry (at half-filling), they have the same sign, the spin-resolved average sign shows similar reentrant behavior, marking the $(\pi,\pi)$-nesting conditions as here. This was seen as an indication that in the thermodynamic limit, with a continuous momentum space, any interaction drives an antiferromagnetic instability. The same logic applied to our case determines that in the Rashba-Hubbard model, any interaction leads to antiferromagnetic order (in the small $U$ regime), resolving the discrepancies in the phase diagrams extracted from mean-field techniques\,\cite{Kennedy2022,Zhang2015,Kawano2023,Kubo2024}.

%%%%%%%%%%%%%%%%%%%%%%%%%%%%%%%%%%%%%%%%%%%%%%%%%%%%%%%%%%%%%%%%%%%%%%%%%%%%%%%%%%%%%
\section{Infinite-$U$ limit}
\label{App:IUL}

\begin{figure}[t]
    \centering
    \includegraphics[scale=0.5]{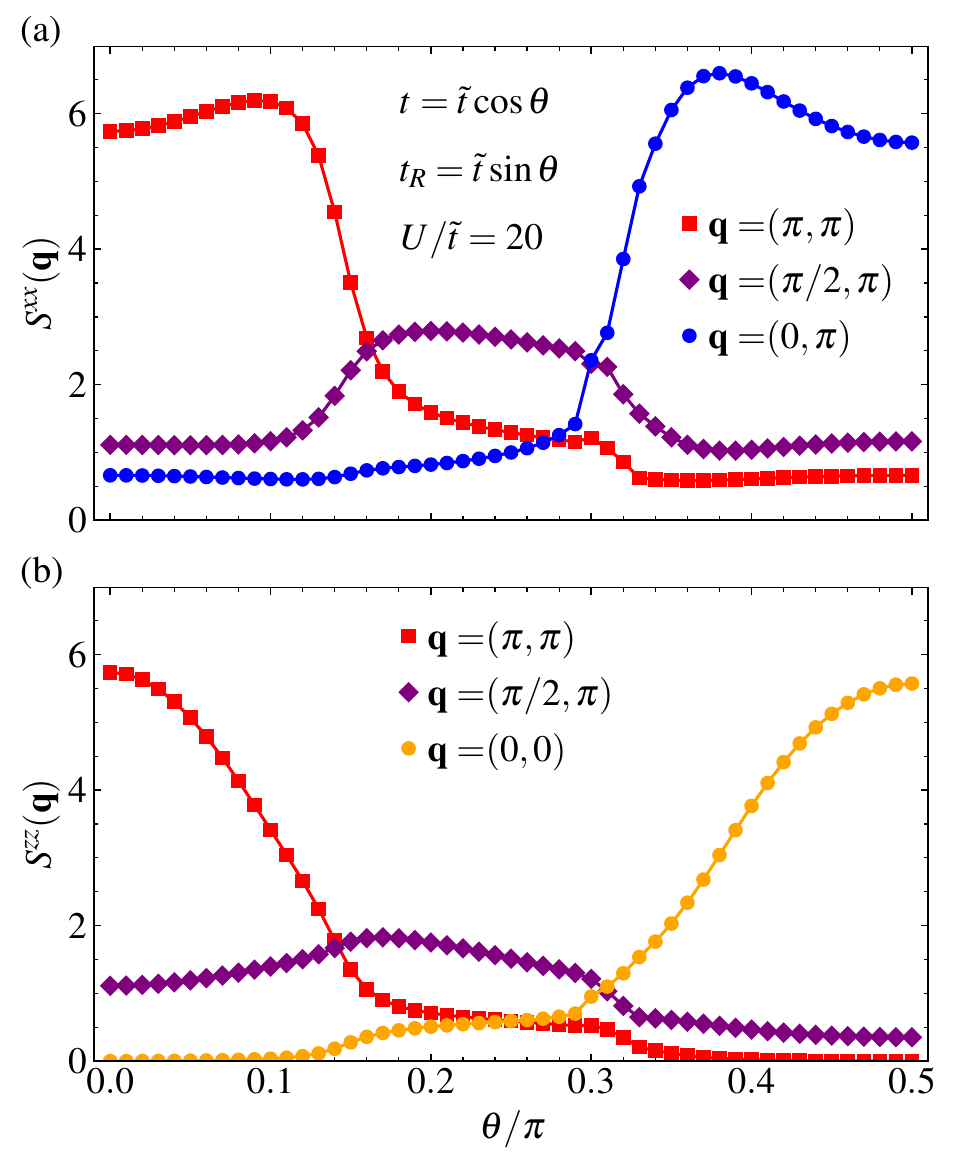}
    \caption{Dominant spin structure factors for (a) $x$ and (b) $z$ spin components. All data are from ED calculations on a $4 \times 4$ lattice with twisted boundary conditions.
    }
    \label{fig:IUL}
\end{figure}

In the presence of the emerging Dzyaloshinskii-Moriya interaction, the $U \to \infty$ limit warrants special attention. In this regime, only the magnetic degrees of freedom remain relevant, and the RHM can be effectively mapped onto a spin Hamiltonian\,\cite{Cole2012,Wang2017,Kawano2023}. Such systems exhibit all the magnetic phases identified in our study. To facilitate a comparison with previous works, we adopt the same hopping parameterization as in Refs.\,\cite{Cole2012,Wang2017}. Specifically, we perform ED calculations with the parameters $t = \tilde{t} \cos{\theta}$, $t_R = \tilde{t} \sin{\theta}$, and $U/\tilde{t} = 20.0$. 

Figure \ref{fig:IUL} illustrates the dependence of the largest structure factors on $\theta$. Using a parameterization similar to the one used in Secs.\,\ref{sec:ED_results} and \ref{sec:DQMC_results}, the RHM displays a ${\bf q} = (\pi,\pi)$ antiferromagnetic phase for $t_R \to 0$, a ${\bf q} = (\pi/2,\pi)$ spiral phase for $t_R \sim t$, and a spin vortex phase as $t \to 0$. Anisotropic magnetic correlations across different spin components characterize these phases. Notably, $S^{xx}(\pi,\pi)$ exhibits a slight increase before decreasing at $\theta/\pi \sim 0.15$, whereas $S^{zz}(\pi,\pi)$ shows a monotonic decline with increasing $\theta$.

In this limit, the RHM also exhibits a $3\times 3$ skyrmion phase\,\cite{Cole2012,Wang2017}, which remains inaccessible on our $4\times 4$ lattice when using ED. Furthermore, the severe sign problem for strong $U$ renders DQMC simulations unfeasible for investigating such phases. 

\section{Spectral properties at the pure Rashba regime.}
\label{App:spectral}

Here, we investigate how the screened Coulomb interactions modify the spectral properties of the Rashba-Hubbard Model (RHM). To this end, we employ cluster perturbation theory (CPT)\,\cite{Senechal2000,Senechal2002,Brosco2020,Fontenele2024}, which enables the study of the lattice Green's function by focusing on smaller tractable clusters. Specifically, we consider a $2 \times 2$ cluster. Within this framework, the Hamiltonian is decomposed as \begin{align} 
\mathcal{\hat H} = \mathcal{\hat H}_{\text{cluster}} + \hat{V}~, \label{eq:Hcpt}
\end{align}
where $\mathcal{\hat H}_{\rm cluster}$ contains the terms acting within the $2 \times 2$ cluster, while $\hat{V}$ accounts for the hopping between different clusters in the superlattice. 

Using exact diagonalization, we compute the cluster Green's function matrix, $\mathbf{G}_{\rm cluster}$. Incorporating the inter-cluster hopping terms via Fourier transform, the Green's function in momentum space can be expressed as
\begin{align} \textbf{G}(\textbf{k},\omega) = \frac{1}{\mathbf{G}_{\text{cluster}}(\omega)^{-1} - \mathbf{V}(\textbf{k})}~.
\label{eq:TFGFcpt2} 
\end{align} 
Here, $\mathbf{V}({\bf k})$ represents the Fourier-transformed hopping matrix, which restores the inter-cluster connectivity.

\begin{figure}[t]
    \centering
    \includegraphics[scale=0.5]{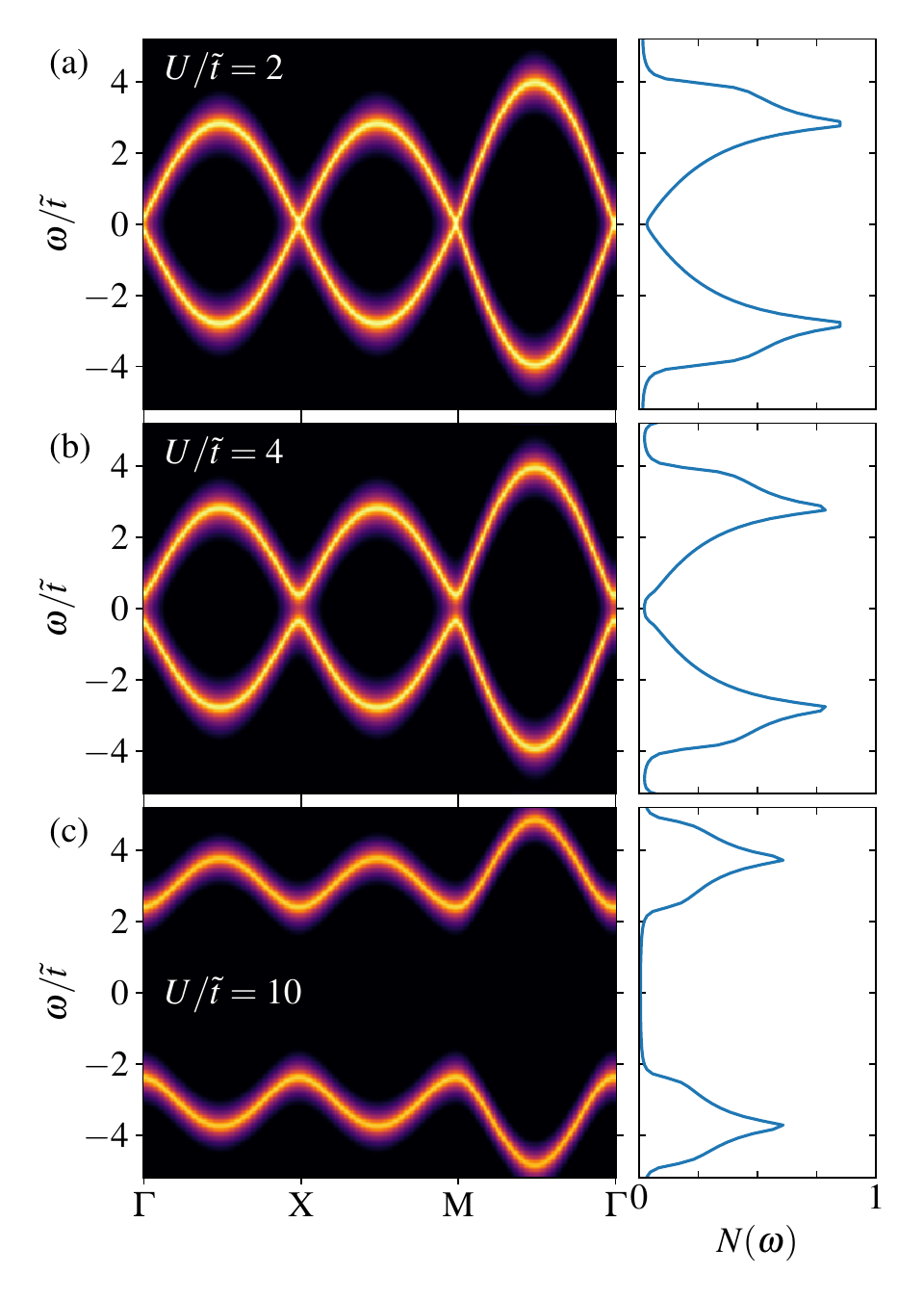}
    \caption{CPT Spectral function and density of states in the pure Rashba case $(\theta = \pi/2)$ for (a) $U/\tilde{t}=2$, (b) $U/\tilde{t}=4$, and (c) $U/\tilde{t}=10$. 
    }
    \label{fig:Akw_dos}
\end{figure}

To recover the periodicity of the full lattice, the CPT Green's function in Fourier space is written as \begin{align} 
G_{\text{CPT}}(\mathbf{k},\omega) = \frac{1}{L}\sum_{a,b}e^{-i\mathbf{k}\cdot(\mathbf{r}_{a} - \mathbf{r}_{b})}G_{a,b}(\mathbf{k},\omega)~, 
\label{eq:GreenCPT} 
\end{align}
where ${\bf r}_a$ and ${\bf r}_b$ are site positions within the cluster. From the CPT Green's functions, we extract the spectral function, 
\begin{align} 
A_{0}(\mathbf{k},\omega) = -\frac{1}{2\pi}\text{Im}[G_{\text{CPT}}(\mathbf{k},\omega)], 
\label{eq:A0} 
\end{align} 
while the density of states is obtained as 
\begin{align} 
N(\omega) = \frac{1}{N}\sum_{\mathbf{k}} A_{0}(\mathbf{k},\omega). 
\label{eq:DOS} 
\end{align}

Figure\,\ref{fig:Akw_dos} illustrates the results for the spectral function and density of states in the pure Rashba case $(\theta = \pi/2)$. The Weyl cones are nearly unaffected for small interactions, as shown in Fig.\,\ref{fig:Akw_dos}(a). The density of states retains a semimetallic profile, with the spectral weight vanishing only at the Fermi level for $U/\tilde{t}=2$. For $U/\tilde{t}=4$, the spectral weight at the Fermi level decreases, indicating a reduction in low-energy states. However, as seen in Fig.\,\ref{fig:Akw_dos}(b), there is still no clear gap in the density of states. For stronger interactions, a gap emerges and becomes evident in both the spectral weight and density of states plots, as displayed in Fig.\,\ref{fig:Akw_dos}(c) for $U/\tilde{t}=10$. This delayed gap opening highlights the resilience of the Weyl semimetallic state at $\theta = \pi/2$ to the emergence of (here, local) spin ordering, contrasting sharply with the $\theta = 0$ case. In the latter scenario, low-energy magnetic excitations induce a robust gap for weaker interactions, even for the small cluster size considered here. 

\begin{figure}[t]
    \centering
    \includegraphics[scale=0.5]{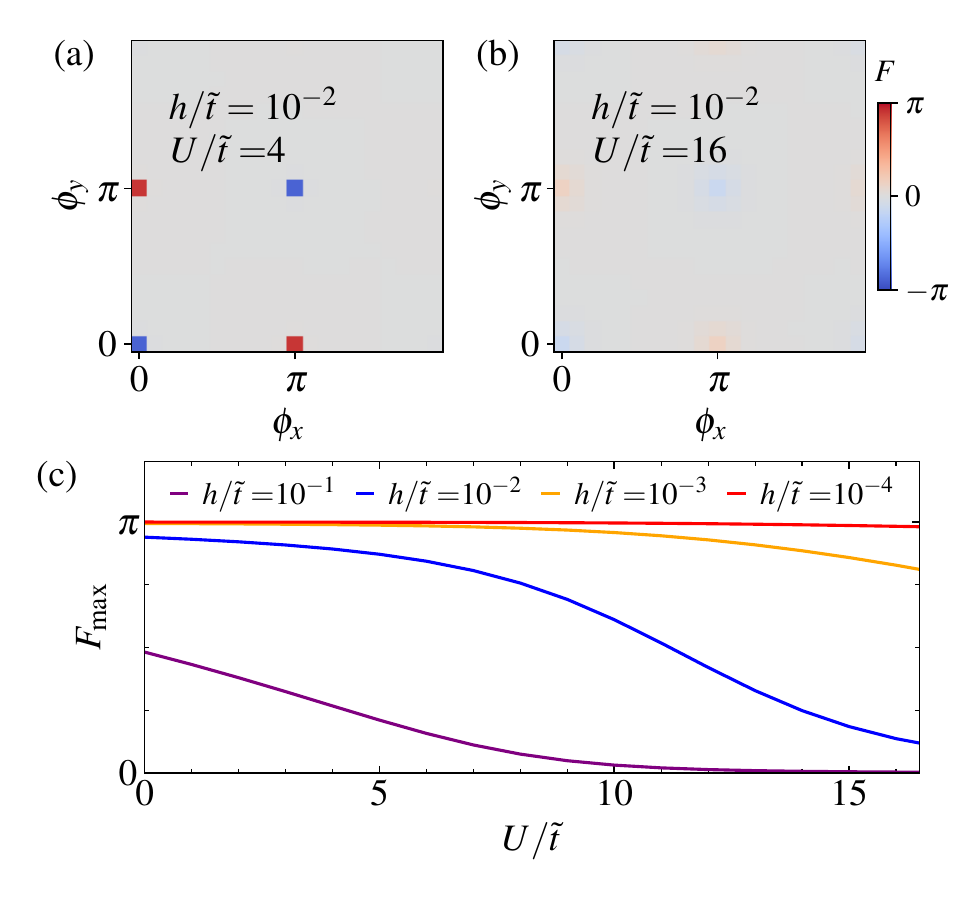}
    \caption{Contour plot of the Berry curvature for a $3 \times 3$ lattice for $\theta = \pi/2$ and small, (a) $U/\tilde t = 4$, and large interactions, (b) $U/\tilde t = 16$. (c) Dependence of the maximum amplitude of the Berry phase singularity on the strength of the Zeeman energy needed to break the spin degeneracy.
    }
    \label{fig:berry_curv}
\end{figure}

\section{Berry Curvature}
\label{App:berry_curv}

In this section, we discuss the topological properties of the RHM in the Weyl semimetallic regime. To this end, we evaluate a discretized version of the Berry curvature in our many-body system using the approach discussed in Refs.\,\cite{Fukui2005,Varney2011}. For that, we compute the many-body ground state in the torus $\{\phi_x,\phi_y\}$ of twisted boundary conditions, where the Berry curvature can be written in terms of the normalized overlaps
\begin{align}
 U^{m,n}_x = \frac{\langle \psi_0^{m,n}|\psi_0^{m+1,n}\rangle}{|\langle \psi_0^{m,n}|\psi_0^{m+1,n}\rangle|}, && U^{m,n}_y = \frac{\langle \psi_0^{m,n}|\psi_0^{m,n+1}\rangle}{|\langle \psi_0^{m,n}|\psi_0^{m,n+1}\rangle|},
\end{align}
at consecutive points of the $\{m,n\}$ grid defined by the $\{\phi_x,\phi_y\}$ phases, when it is subdivided in $N_\alpha$ intervals, such as $\phi_\alpha = 2\pi i/N_\alpha$. Thus, the discretized Berry curvature assumes the form
\begin{equation}
  F_{m,n} = -{i}\log\left(\frac{U_x^{m,n} U_y^{m+1,n}}{U_x^{m,n+1}U_y^{m,n}}\right),
\end{equation}
where an approximant of the Chern number, $C = (1/2\pi)\sum_{m,n} F_{m,n}$, converges to the true value when $N_\alpha \to \infty$. Figure \ref{fig:berry_curv} summarizes the results for the Berry curvature for a $3\times 3$ lattice, where we use $N_\alpha = 20$. We notice that the calculation of the Berry curvature above assumes a finite gap in the many-body spectrum manifold\,\cite{Fukui2005}. To achieve that, we introduce a small perturbation, breaking the degeneracy of the two spin energy bands by introducing a Zeeman field term into the Hamiltonian, $\mathcal{\hat H}_Z = -h \sum_\iv (\hat n_{\iv,\uparrow} - \hat n_{\iv,\downarrow})$. As a result, the Berry curvature in Fig.\,\ref{fig:berry_curv} displays the expected behavior for a two-dimensional Weyl semi-metal for finite, but small, interactions: two pairs of singularities with opposite chiralities, such that $C=0$. When increasing the interaction strength, these singularities are suppressed, resulting in a topologically trivial phase, which we have identified in the main text with one exhibiting a local order parameter describing a spin-vortex phase. Note that this analysis is merely qualitative owing to the very small lattice used.

%%%%%%%%%%%%%%%%%%%%%%%%%%%%%%%%%%%%%%%%%%%%%%%%%%%%%%%%%%%%%%%%%%%%%%%      Refs          %%%%%%%%%%%%%%%%%%%
%%%%%%%%%%%%%%%%%%%%%%%%%%%%%%%%%%%%%%%%%%%%%%%%%%%%%%%
\bibliography{ref}

\begin{thebibliography}{75}%
\makeatletter
\providecommand \@ifxundefined [1]{%
 \@ifx{#1\undefined}
}%
\providecommand \@ifnum [1]{%
 \ifnum #1\expandafter \@firstoftwo
 \else \expandafter \@secondoftwo
 \fi
}%
\providecommand \@ifx [1]{%
 \ifx #1\expandafter \@firstoftwo
 \else \expandafter \@secondoftwo
 \fi
}%
\providecommand \natexlab [1]{#1}%
\providecommand \enquote  [1]{``#1''}%
\providecommand \bibnamefont  [1]{#1}%
\providecommand \bibfnamefont [1]{#1}%
\providecommand \citenamefont [1]{#1}%
\providecommand \href@noop [0]{\@secondoftwo}%
\providecommand \href [0]{\begingroup \@sanitize@url \@href}%
\providecommand \@href[1]{\@@startlink{#1}\@@href}%
\providecommand \@@href[1]{\endgroup#1\@@endlink}%
\providecommand \@sanitize@url [0]{\catcode `\\12\catcode `\$12\catcode `\&12\catcode `\#12\catcode `\^12\catcode `\_12\catcode `\%12\relax}%
\providecommand \@@startlink[1]{}%
\providecommand \@@endlink[0]{}%
\providecommand \url  [0]{\begingroup\@sanitize@url \@url }%
\providecommand \@url [1]{\endgroup\@href {#1}{\urlprefix }}%
\providecommand \urlprefix  [0]{URL }%
\providecommand \Eprint [0]{\href }%
\providecommand \doibase [0]{https://doi.org/}%
\providecommand \selectlanguage [0]{\@gobble}%
\providecommand \bibinfo  [0]{\@secondoftwo}%
\providecommand \bibfield  [0]{\@secondoftwo}%
\providecommand \translation [1]{[#1]}%
\providecommand \BibitemOpen [0]{}%
\providecommand \bibitemStop [0]{}%
\providecommand \bibitemNoStop [0]{.\EOS\space}%
\providecommand \EOS [0]{\spacefactor3000\relax}%
\providecommand \BibitemShut  [1]{\csname bibitem#1\endcsname}%
\let\auto@bib@innerbib\@empty
%</preamble>
\bibitem [{\citenamefont {Nitta}\ \emph {et~al.}(1997)\citenamefont {Nitta}, \citenamefont {Akazaki}, \citenamefont {Takayanagi},\ and\ \citenamefont {Enoki}}]{Nitta1997}%
  \BibitemOpen
  \bibfield  {author} {\bibinfo {author} {\bibfnamefont {J.}~\bibnamefont {Nitta}}, \bibinfo {author} {\bibfnamefont {T.}~\bibnamefont {Akazaki}}, \bibinfo {author} {\bibfnamefont {H.}~\bibnamefont {Takayanagi}},\ and\ \bibinfo {author} {\bibfnamefont {T.}~\bibnamefont {Enoki}},\ }\bibfield  {title} {\bibinfo {title} {{Gate Control of Spin-Orbit Interaction in an Inverted I${\mathrm{n}}_{0.53}$G${\mathrm{a}}_{0.47}$As/I${\mathrm{n}}_{0.52}$A${\mathrm{l}}_{0.48}$As Heterostructure}},\ }\href {https://doi.org/10.1103/PhysRevLett.78.1335} {\bibfield  {journal} {\bibinfo  {journal} {Phys. Rev. Lett.}\ }\textbf {\bibinfo {volume} {78}},\ \bibinfo {pages} {1335} (\bibinfo {year} {1997})}\BibitemShut {NoStop}%
\bibitem [{\citenamefont {Caviglia}\ \emph {et~al.}(2010)\citenamefont {Caviglia}, \citenamefont {Gabay}, \citenamefont {Gariglio}, \citenamefont {Reyren}, \citenamefont {Cancellieri},\ and\ \citenamefont {Triscone}}]{Caviglia2010}%
  \BibitemOpen
  \bibfield  {author} {\bibinfo {author} {\bibfnamefont {A.~D.}\ \bibnamefont {Caviglia}}, \bibinfo {author} {\bibfnamefont {M.}~\bibnamefont {Gabay}}, \bibinfo {author} {\bibfnamefont {S.}~\bibnamefont {Gariglio}}, \bibinfo {author} {\bibfnamefont {N.}~\bibnamefont {Reyren}}, \bibinfo {author} {\bibfnamefont {C.}~\bibnamefont {Cancellieri}},\ and\ \bibinfo {author} {\bibfnamefont {J.-M.}\ \bibnamefont {Triscone}},\ }\bibfield  {title} {\bibinfo {title} {Tunable {R}ashba spin-orbit interaction at oxide interfaces},\ }\href {https://doi.org/10.1103/PhysRevLett.104.126803} {\bibfield  {journal} {\bibinfo  {journal} {Phys. Rev. Lett.}\ }\textbf {\bibinfo {volume} {104}},\ \bibinfo {pages} {126803} (\bibinfo {year} {2010})}\BibitemShut {NoStop}%
\bibitem [{\citenamefont {LaShell}\ \emph {et~al.}(1996)\citenamefont {LaShell}, \citenamefont {McDougall},\ and\ \citenamefont {Jensen}}]{LaShell1996}%
  \BibitemOpen
  \bibfield  {author} {\bibinfo {author} {\bibfnamefont {S.}~\bibnamefont {LaShell}}, \bibinfo {author} {\bibfnamefont {B.~A.}\ \bibnamefont {McDougall}},\ and\ \bibinfo {author} {\bibfnamefont {E.}~\bibnamefont {Jensen}},\ }\bibfield  {title} {\bibinfo {title} {{Spin Splitting of an Au(111) Surface State Band Observed with Angle Resolved Photoelectron Spectroscopy}},\ }\href {https://doi.org/10.1103/PhysRevLett.77.3419} {\bibfield  {journal} {\bibinfo  {journal} {Phys. Rev. Lett.}\ }\textbf {\bibinfo {volume} {77}},\ \bibinfo {pages} {3419} (\bibinfo {year} {1996})}\BibitemShut {NoStop}%
\bibitem [{\citenamefont {Sunko}\ \emph {et~al.}(2017)\citenamefont {Sunko}, \citenamefont {Rosner}, \citenamefont {Kushwaha}, \citenamefont {Khim}, \citenamefont {Mazzola}, \citenamefont {Bawden}, \citenamefont {Clark}, \citenamefont {Riley}, \citenamefont {Kasinathan}, \citenamefont {Haverkort}, \citenamefont {Kim}, \citenamefont {Hoesch}, \citenamefont {Fujii}, \citenamefont {Vobornik}, \citenamefont {Mackenzie},\ and\ \citenamefont {King}}]{Sunko2017}%
  \BibitemOpen
  \bibfield  {author} {\bibinfo {author} {\bibfnamefont {V.}~\bibnamefont {Sunko}}, \bibinfo {author} {\bibfnamefont {H.}~\bibnamefont {Rosner}}, \bibinfo {author} {\bibfnamefont {P.}~\bibnamefont {Kushwaha}}, \bibinfo {author} {\bibfnamefont {S.}~\bibnamefont {Khim}}, \bibinfo {author} {\bibfnamefont {F.}~\bibnamefont {Mazzola}}, \bibinfo {author} {\bibfnamefont {L.}~\bibnamefont {Bawden}}, \bibinfo {author} {\bibfnamefont {O.~J.}\ \bibnamefont {Clark}}, \bibinfo {author} {\bibfnamefont {J.~M.}\ \bibnamefont {Riley}}, \bibinfo {author} {\bibfnamefont {D.}~\bibnamefont {Kasinathan}}, \bibinfo {author} {\bibfnamefont {M.~W.}\ \bibnamefont {Haverkort}}, \bibinfo {author} {\bibfnamefont {T.~K.}\ \bibnamefont {Kim}}, \bibinfo {author} {\bibfnamefont {M.}~\bibnamefont {Hoesch}}, \bibinfo {author} {\bibfnamefont {J.}~\bibnamefont {Fujii}}, \bibinfo {author} {\bibfnamefont {I.}~\bibnamefont {Vobornik}}, \bibinfo {author} {\bibfnamefont {A.~P.}\ \bibnamefont {Mackenzie}},\ and\ \bibinfo {author} {\bibfnamefont
  {P.~D.~C.}\ \bibnamefont {King}},\ }\bibfield  {title} {\bibinfo {title} {Maximal {R}ashba-like spin splitting via kinetic-energy-coupled inversion-symmetry breaking},\ }\href {https://doi.org/10.1038/nature23898} {\bibfield  {journal} {\bibinfo  {journal} {Nature}\ }\textbf {\bibinfo {volume} {549}},\ \bibinfo {pages} {492} (\bibinfo {year} {2017})}\BibitemShut {NoStop}%
\bibitem [{\citenamefont {Min}\ \emph {et~al.}(2006)\citenamefont {Min}, \citenamefont {Hill}, \citenamefont {Sinitsyn}, \citenamefont {Sahu}, \citenamefont {Kleinman},\ and\ \citenamefont {MacDonald}}]{Min2006}%
  \BibitemOpen
  \bibfield  {author} {\bibinfo {author} {\bibfnamefont {H.}~\bibnamefont {Min}}, \bibinfo {author} {\bibfnamefont {J.~E.}\ \bibnamefont {Hill}}, \bibinfo {author} {\bibfnamefont {N.~A.}\ \bibnamefont {Sinitsyn}}, \bibinfo {author} {\bibfnamefont {B.~R.}\ \bibnamefont {Sahu}}, \bibinfo {author} {\bibfnamefont {L.}~\bibnamefont {Kleinman}},\ and\ \bibinfo {author} {\bibfnamefont {A.~H.}\ \bibnamefont {MacDonald}},\ }\bibfield  {title} {\bibinfo {title} {{Intrinsic and Rashba spin-orbit interactions in graphene sheets}},\ }\href {https://doi.org/10.1103/PhysRevB.74.165310} {\bibfield  {journal} {\bibinfo  {journal} {Phys. Rev. B}\ }\textbf {\bibinfo {volume} {74}},\ \bibinfo {pages} {165310} (\bibinfo {year} {2006})}\BibitemShut {NoStop}%
\bibitem [{\citenamefont {Manchon}\ \emph {et~al.}(2015)\citenamefont {Manchon}, \citenamefont {Koo}, \citenamefont {Nitta}, \citenamefont {Frolov},\ and\ \citenamefont {Duine}}]{Manchon2015}%
  \BibitemOpen
  \bibfield  {author} {\bibinfo {author} {\bibfnamefont {A.}~\bibnamefont {Manchon}}, \bibinfo {author} {\bibfnamefont {H.~C.}\ \bibnamefont {Koo}}, \bibinfo {author} {\bibfnamefont {J.}~\bibnamefont {Nitta}}, \bibinfo {author} {\bibfnamefont {S.~M.}\ \bibnamefont {Frolov}},\ and\ \bibinfo {author} {\bibfnamefont {R.~A.}\ \bibnamefont {Duine}},\ }\bibfield  {title} {\bibinfo {title} {New perspectives for {R}ashba spin--orbit coupling},\ }\href {https://doi.org/10.1038/nmat4360} {\bibfield  {journal} {\bibinfo  {journal} {Nature Materials}\ }\textbf {\bibinfo {volume} {14}},\ \bibinfo {pages} {871} (\bibinfo {year} {2015})}\BibitemShut {NoStop}%
\bibitem [{\citenamefont {Schultz}\ \emph {et~al.}(1996)\citenamefont {Schultz}, \citenamefont {Heinrichs}, \citenamefont {Merkt}, \citenamefont {Colin}, \citenamefont {Skauli},\ and\ \citenamefont {Løvold}}]{Schultz1996}%
  \BibitemOpen
  \bibfield  {author} {\bibinfo {author} {\bibfnamefont {M.}~\bibnamefont {Schultz}}, \bibinfo {author} {\bibfnamefont {F.}~\bibnamefont {Heinrichs}}, \bibinfo {author} {\bibfnamefont {U.}~\bibnamefont {Merkt}}, \bibinfo {author} {\bibfnamefont {T.}~\bibnamefont {Colin}}, \bibinfo {author} {\bibfnamefont {T.}~\bibnamefont {Skauli}},\ and\ \bibinfo {author} {\bibfnamefont {S.}~\bibnamefont {Løvold}},\ }\bibfield  {title} {\bibinfo {title} {Rashba spin splitting in a gated {HgTe} quantum well},\ }\href {https://doi.org/10.1088/0268-1242/11/8/009} {\bibfield  {journal} {\bibinfo  {journal} {Semiconductor Science and Technology}\ }\textbf {\bibinfo {volume} {11}},\ \bibinfo {pages} {1168} (\bibinfo {year} {1996})}\BibitemShut {NoStop}%
\bibitem [{\citenamefont {King}\ \emph {et~al.}(2011)\citenamefont {King}, \citenamefont {Hatch}, \citenamefont {Bianchi}, \citenamefont {Ovsyannikov}, \citenamefont {Lupulescu}, \citenamefont {Landolt}, \citenamefont {Slomski}, \citenamefont {Dil}, \citenamefont {Guan}, \citenamefont {Mi}, \citenamefont {Rienks}, \citenamefont {Fink}, \citenamefont {Lindblad}, \citenamefont {Svensson}, \citenamefont {Bao}, \citenamefont {Balakrishnan}, \citenamefont {Iversen}, \citenamefont {Osterwalder}, \citenamefont {Eberhardt}, \citenamefont {Baumberger},\ and\ \citenamefont {Hofmann}}]{King2011}%
  \BibitemOpen
  \bibfield  {author} {\bibinfo {author} {\bibfnamefont {P.~D.~C.}\ \bibnamefont {King}}, \bibinfo {author} {\bibfnamefont {R.~C.}\ \bibnamefont {Hatch}}, \bibinfo {author} {\bibfnamefont {M.}~\bibnamefont {Bianchi}}, \bibinfo {author} {\bibfnamefont {R.}~\bibnamefont {Ovsyannikov}}, \bibinfo {author} {\bibfnamefont {C.}~\bibnamefont {Lupulescu}}, \bibinfo {author} {\bibfnamefont {G.}~\bibnamefont {Landolt}}, \bibinfo {author} {\bibfnamefont {B.}~\bibnamefont {Slomski}}, \bibinfo {author} {\bibfnamefont {J.~H.}\ \bibnamefont {Dil}}, \bibinfo {author} {\bibfnamefont {D.}~\bibnamefont {Guan}}, \bibinfo {author} {\bibfnamefont {J.~L.}\ \bibnamefont {Mi}}, \bibinfo {author} {\bibfnamefont {E.~D.~L.}\ \bibnamefont {Rienks}}, \bibinfo {author} {\bibfnamefont {J.}~\bibnamefont {Fink}}, \bibinfo {author} {\bibfnamefont {A.}~\bibnamefont {Lindblad}}, \bibinfo {author} {\bibfnamefont {S.}~\bibnamefont {Svensson}}, \bibinfo {author} {\bibfnamefont {S.}~\bibnamefont {Bao}}, \bibinfo {author} {\bibfnamefont
  {G.}~\bibnamefont {Balakrishnan}}, \bibinfo {author} {\bibfnamefont {B.~B.}\ \bibnamefont {Iversen}}, \bibinfo {author} {\bibfnamefont {J.}~\bibnamefont {Osterwalder}}, \bibinfo {author} {\bibfnamefont {W.}~\bibnamefont {Eberhardt}}, \bibinfo {author} {\bibfnamefont {F.}~\bibnamefont {Baumberger}},\ and\ \bibinfo {author} {\bibfnamefont {P.}~\bibnamefont {Hofmann}},\ }\bibfield  {title} {\bibinfo {title} {{Large Tunable Rashba Spin Splitting of a Two-Dimensional Electron Gas in ${\mathrm{Bi}}_{2}{\mathrm{Se}}_{3}$}},\ }\href {https://doi.org/10.1103/PhysRevLett.107.096802} {\bibfield  {journal} {\bibinfo  {journal} {Phys. Rev. Lett.}\ }\textbf {\bibinfo {volume} {107}},\ \bibinfo {pages} {096802} (\bibinfo {year} {2011})}\BibitemShut {NoStop}%
\bibitem [{\citenamefont {Datta}\ and\ \citenamefont {Das}(1990)}]{Datta1990}%
  \BibitemOpen
  \bibfield  {author} {\bibinfo {author} {\bibfnamefont {S.}~\bibnamefont {Datta}}\ and\ \bibinfo {author} {\bibfnamefont {B.}~\bibnamefont {Das}},\ }\bibfield  {title} {\bibinfo {title} {{Electronic analog of the electro‐optic modulator}},\ }\href {https://doi.org/10.1063/1.102730} {\bibfield  {journal} {\bibinfo  {journal} {Applied Physics Letters}\ }\textbf {\bibinfo {volume} {56}},\ \bibinfo {pages} {665} (\bibinfo {year} {1990})}\BibitemShut {NoStop}%
\bibitem [{\citenamefont {Koo}\ \emph {et~al.}(2009)\citenamefont {Koo}, \citenamefont {Kwon}, \citenamefont {Eom}, \citenamefont {Chang}, \citenamefont {Han},\ and\ \citenamefont {Johnson}}]{Koo2009}%
  \BibitemOpen
  \bibfield  {author} {\bibinfo {author} {\bibfnamefont {H.~C.}\ \bibnamefont {Koo}}, \bibinfo {author} {\bibfnamefont {J.~H.}\ \bibnamefont {Kwon}}, \bibinfo {author} {\bibfnamefont {J.}~\bibnamefont {Eom}}, \bibinfo {author} {\bibfnamefont {J.}~\bibnamefont {Chang}}, \bibinfo {author} {\bibfnamefont {S.~H.}\ \bibnamefont {Han}},\ and\ \bibinfo {author} {\bibfnamefont {M.}~\bibnamefont {Johnson}},\ }\bibfield  {title} {\bibinfo {title} {Control of spin precession in a spin-injected field effect transistor},\ }\href {https://doi.org/10.1126/science.1173667} {\bibfield  {journal} {\bibinfo  {journal} {Science}\ }\textbf {\bibinfo {volume} {325}},\ \bibinfo {pages} {1515} (\bibinfo {year} {2009})}\BibitemShut {NoStop}%
\bibitem [{\citenamefont {Murakami}\ \emph {et~al.}(2003)\citenamefont {Murakami}, \citenamefont {Nagaosa},\ and\ \citenamefont {Zhang}}]{Murakami2003}%
  \BibitemOpen
  \bibfield  {author} {\bibinfo {author} {\bibfnamefont {S.}~\bibnamefont {Murakami}}, \bibinfo {author} {\bibfnamefont {N.}~\bibnamefont {Nagaosa}},\ and\ \bibinfo {author} {\bibfnamefont {S.-C.}\ \bibnamefont {Zhang}},\ }\bibfield  {title} {\bibinfo {title} {Dissipationless quantum spin current at room temperature},\ }\href {https://doi.org/10.1126/science.1087128} {\bibfield  {journal} {\bibinfo  {journal} {Science}\ }\textbf {\bibinfo {volume} {301}},\ \bibinfo {pages} {1348} (\bibinfo {year} {2003})}\BibitemShut {NoStop}%
\bibitem [{\citenamefont {Sinova}\ \emph {et~al.}(2004)\citenamefont {Sinova}, \citenamefont {Culcer}, \citenamefont {Niu}, \citenamefont {Sinitsyn}, \citenamefont {Jungwirth},\ and\ \citenamefont {MacDonald}}]{Sinova2004}%
  \BibitemOpen
  \bibfield  {author} {\bibinfo {author} {\bibfnamefont {J.}~\bibnamefont {Sinova}}, \bibinfo {author} {\bibfnamefont {D.}~\bibnamefont {Culcer}}, \bibinfo {author} {\bibfnamefont {Q.}~\bibnamefont {Niu}}, \bibinfo {author} {\bibfnamefont {N.~A.}\ \bibnamefont {Sinitsyn}}, \bibinfo {author} {\bibfnamefont {T.}~\bibnamefont {Jungwirth}},\ and\ \bibinfo {author} {\bibfnamefont {A.~H.}\ \bibnamefont {MacDonald}},\ }\bibfield  {title} {\bibinfo {title} {{Universal Intrinsic Spin Hall Effect}},\ }\href {https://doi.org/10.1103/PhysRevLett.92.126603} {\bibfield  {journal} {\bibinfo  {journal} {Phys. Rev. Lett.}\ }\textbf {\bibinfo {volume} {92}},\ \bibinfo {pages} {126603} (\bibinfo {year} {2004})}\BibitemShut {NoStop}%
\bibitem [{\citenamefont {Dzyaloshinsky}(1958)}]{Dzyaloshinsky1958}%
  \BibitemOpen
  \bibfield  {author} {\bibinfo {author} {\bibfnamefont {I.}~\bibnamefont {Dzyaloshinsky}},\ }\bibfield  {title} {\bibinfo {title} {{A thermodynamic theory of “weak” ferromagnetism of antiferromagnetics}},\ }\href {https://doi.org/https://doi.org/10.1016/0022-3697(58)90076-3} {\bibfield  {journal} {\bibinfo  {journal} {Journal of Physics and Chemistry of Solids}\ }\textbf {\bibinfo {volume} {4}},\ \bibinfo {pages} {241} (\bibinfo {year} {1958})}\BibitemShut {NoStop}%
\bibitem [{\citenamefont {Moriya}(1960)}]{Moriya1960}%
  \BibitemOpen
  \bibfield  {author} {\bibinfo {author} {\bibfnamefont {T.}~\bibnamefont {Moriya}},\ }\bibfield  {title} {\bibinfo {title} {{Anisotropic Superexchange Interaction and Weak Ferromagnetism}},\ }\href {https://doi.org/10.1103/PhysRev.120.91} {\bibfield  {journal} {\bibinfo  {journal} {Phys. Rev.}\ }\textbf {\bibinfo {volume} {120}},\ \bibinfo {pages} {91} (\bibinfo {year} {1960})}\BibitemShut {NoStop}%
\bibitem [{\citenamefont {Sergienko}\ and\ \citenamefont {Dagotto}(2006)}]{Sergienko2006}%
  \BibitemOpen
  \bibfield  {author} {\bibinfo {author} {\bibfnamefont {I.~A.}\ \bibnamefont {Sergienko}}\ and\ \bibinfo {author} {\bibfnamefont {E.}~\bibnamefont {Dagotto}},\ }\bibfield  {title} {\bibinfo {title} {{Role of the Dzyaloshinskii-Moriya interaction in multiferroic perovskites}},\ }\href {https://doi.org/10.1103/PhysRevB.73.094434} {\bibfield  {journal} {\bibinfo  {journal} {Phys. Rev. B}\ }\textbf {\bibinfo {volume} {73}},\ \bibinfo {pages} {094434} (\bibinfo {year} {2006})}\BibitemShut {NoStop}%
\bibitem [{\citenamefont {Seki}\ \emph {et~al.}(2012)\citenamefont {Seki}, \citenamefont {Yu}, \citenamefont {Ishiwata},\ and\ \citenamefont {Tokura}}]{Seki2012}%
  \BibitemOpen
  \bibfield  {author} {\bibinfo {author} {\bibfnamefont {S.}~\bibnamefont {Seki}}, \bibinfo {author} {\bibfnamefont {X.~Z.}\ \bibnamefont {Yu}}, \bibinfo {author} {\bibfnamefont {S.}~\bibnamefont {Ishiwata}},\ and\ \bibinfo {author} {\bibfnamefont {Y.}~\bibnamefont {Tokura}},\ }\bibfield  {title} {\bibinfo {title} {{Observation of Skyrmions in a Multiferroic Material}},\ }\href {https://doi.org/10.1126/science.1214143} {\bibfield  {journal} {\bibinfo  {journal} {Science}\ }\textbf {\bibinfo {volume} {336}},\ \bibinfo {pages} {198} (\bibinfo {year} {2012})}\BibitemShut {NoStop}%
\bibitem [{\citenamefont {Nagaosa}\ and\ \citenamefont {Tokura}(2013)}]{Nagaosa2013}%
  \BibitemOpen
  \bibfield  {author} {\bibinfo {author} {\bibfnamefont {N.}~\bibnamefont {Nagaosa}}\ and\ \bibinfo {author} {\bibfnamefont {Y.}~\bibnamefont {Tokura}},\ }\bibfield  {title} {\bibinfo {title} {{Topological properties and dynamics of magnetic skyrmions}},\ }\href {https://doi.org/10.1038/nnano.2013.243} {\bibfield  {journal} {\bibinfo  {journal} {Nature Nanotechnology}\ }\textbf {\bibinfo {volume} {8}},\ \bibinfo {pages} {899} (\bibinfo {year} {2013})}\BibitemShut {NoStop}%
\bibitem [{\citenamefont {Wilson}\ \emph {et~al.}(2014)\citenamefont {Wilson}, \citenamefont {Butenko}, \citenamefont {Bogdanov},\ and\ \citenamefont {Monchesky}}]{Wilson2014}%
  \BibitemOpen
  \bibfield  {author} {\bibinfo {author} {\bibfnamefont {M.~N.}\ \bibnamefont {Wilson}}, \bibinfo {author} {\bibfnamefont {A.~B.}\ \bibnamefont {Butenko}}, \bibinfo {author} {\bibfnamefont {A.~N.}\ \bibnamefont {Bogdanov}},\ and\ \bibinfo {author} {\bibfnamefont {T.~L.}\ \bibnamefont {Monchesky}},\ }\bibfield  {title} {\bibinfo {title} {{Chiral skyrmions in cubic helimagnet films: The role of uniaxial anisotropy}},\ }\href {https://doi.org/10.1103/PhysRevB.89.094411} {\bibfield  {journal} {\bibinfo  {journal} {Phys. Rev. B}\ }\textbf {\bibinfo {volume} {89}},\ \bibinfo {pages} {094411} (\bibinfo {year} {2014})}\BibitemShut {NoStop}%
\bibitem [{\citenamefont {Yang}\ \emph {et~al.}(2018)\citenamefont {Yang}, \citenamefont {Chen}, \citenamefont {Cotta}, \citenamefont {N'Diaye}, \citenamefont {Nikolaev}, \citenamefont {Soares}, \citenamefont {Macedo}, \citenamefont {Liu}, \citenamefont {Schmid}, \citenamefont {Fert},\ and\ \citenamefont {Chshiev}}]{Yang2018}%
  \BibitemOpen
  \bibfield  {author} {\bibinfo {author} {\bibfnamefont {H.}~\bibnamefont {Yang}}, \bibinfo {author} {\bibfnamefont {G.}~\bibnamefont {Chen}}, \bibinfo {author} {\bibfnamefont {A.~A.~C.}\ \bibnamefont {Cotta}}, \bibinfo {author} {\bibfnamefont {A.~T.}\ \bibnamefont {N'Diaye}}, \bibinfo {author} {\bibfnamefont {S.~A.}\ \bibnamefont {Nikolaev}}, \bibinfo {author} {\bibfnamefont {E.~A.}\ \bibnamefont {Soares}}, \bibinfo {author} {\bibfnamefont {W.~A.~A.}\ \bibnamefont {Macedo}}, \bibinfo {author} {\bibfnamefont {K.}~\bibnamefont {Liu}}, \bibinfo {author} {\bibfnamefont {A.~K.}\ \bibnamefont {Schmid}}, \bibinfo {author} {\bibfnamefont {A.}~\bibnamefont {Fert}},\ and\ \bibinfo {author} {\bibfnamefont {M.}~\bibnamefont {Chshiev}},\ }\bibfield  {title} {\bibinfo {title} {Significant {D}zyaloshinskii--{M}oriya interaction at graphene--ferromagnet interfaces due to the {R}ashba effect},\ }\href {https://doi.org/10.1038/s41563-018-0079-4} {\bibfield  {journal} {\bibinfo  {journal} {Nature Materials}\ }\textbf {\bibinfo
  {volume} {17}},\ \bibinfo {pages} {605} (\bibinfo {year} {2018})}\BibitemShut {NoStop}%
\bibitem [{\citenamefont {Hallal}\ \emph {et~al.}(2021)\citenamefont {Hallal}, \citenamefont {Liang}, \citenamefont {Ibrahim}, \citenamefont {Yang}, \citenamefont {Fert},\ and\ \citenamefont {Chshiev}}]{Hallal2021}%
  \BibitemOpen
  \bibfield  {author} {\bibinfo {author} {\bibfnamefont {A.}~\bibnamefont {Hallal}}, \bibinfo {author} {\bibfnamefont {J.}~\bibnamefont {Liang}}, \bibinfo {author} {\bibfnamefont {F.}~\bibnamefont {Ibrahim}}, \bibinfo {author} {\bibfnamefont {H.}~\bibnamefont {Yang}}, \bibinfo {author} {\bibfnamefont {A.}~\bibnamefont {Fert}},\ and\ \bibinfo {author} {\bibfnamefont {M.}~\bibnamefont {Chshiev}},\ }\bibfield  {title} {\bibinfo {title} {Rashba-type {D}zyaloshinskii–{M}oriya interaction, perpendicular magnetic anisotropy, and skyrmion states at 2{D} materials/{C}o interfaces},\ }\href {https://doi.org/10.1021/acs.nanolett.1c01713} {\bibfield  {journal} {\bibinfo  {journal} {Nano Letters}\ }\textbf {\bibinfo {volume} {21}},\ \bibinfo {pages} {7138} (\bibinfo {year} {2021})}\BibitemShut {NoStop}%
\bibitem [{\citenamefont {Cheuk}\ \emph {et~al.}(2012)\citenamefont {Cheuk}, \citenamefont {Sommer}, \citenamefont {Hadzibabic}, \citenamefont {Yefsah}, \citenamefont {Bakr},\ and\ \citenamefont {Zwierlein}}]{Cheuk2012}%
  \BibitemOpen
  \bibfield  {author} {\bibinfo {author} {\bibfnamefont {L.~W.}\ \bibnamefont {Cheuk}}, \bibinfo {author} {\bibfnamefont {A.~T.}\ \bibnamefont {Sommer}}, \bibinfo {author} {\bibfnamefont {Z.}~\bibnamefont {Hadzibabic}}, \bibinfo {author} {\bibfnamefont {T.}~\bibnamefont {Yefsah}}, \bibinfo {author} {\bibfnamefont {W.~S.}\ \bibnamefont {Bakr}},\ and\ \bibinfo {author} {\bibfnamefont {M.~W.}\ \bibnamefont {Zwierlein}},\ }\bibfield  {title} {\bibinfo {title} {Spin-injection spectroscopy of a spin-orbit coupled {F}ermi gas},\ }\href {https://doi.org/10.1103/PhysRevLett.109.095302} {\bibfield  {journal} {\bibinfo  {journal} {Phys. Rev. Lett.}\ }\textbf {\bibinfo {volume} {109}},\ \bibinfo {pages} {095302} (\bibinfo {year} {2012})}\BibitemShut {NoStop}%
\bibitem [{\citenamefont {Ji}\ \emph {et~al.}(2014)\citenamefont {Ji}, \citenamefont {Zhang}, \citenamefont {Zhang}, \citenamefont {Du}, \citenamefont {Zheng}, \citenamefont {Deng}, \citenamefont {Zhai}, \citenamefont {Chen},\ and\ \citenamefont {Pan}}]{Ji2014}%
  \BibitemOpen
  \bibfield  {author} {\bibinfo {author} {\bibfnamefont {S.-C.}\ \bibnamefont {Ji}}, \bibinfo {author} {\bibfnamefont {J.-Y.}\ \bibnamefont {Zhang}}, \bibinfo {author} {\bibfnamefont {L.}~\bibnamefont {Zhang}}, \bibinfo {author} {\bibfnamefont {Z.-D.}\ \bibnamefont {Du}}, \bibinfo {author} {\bibfnamefont {W.}~\bibnamefont {Zheng}}, \bibinfo {author} {\bibfnamefont {Y.-J.}\ \bibnamefont {Deng}}, \bibinfo {author} {\bibfnamefont {H.}~\bibnamefont {Zhai}}, \bibinfo {author} {\bibfnamefont {S.}~\bibnamefont {Chen}},\ and\ \bibinfo {author} {\bibfnamefont {J.-W.}\ \bibnamefont {Pan}},\ }\bibfield  {title} {\bibinfo {title} {Experimental determination of the finite-temperature phase diagram of a spin--orbit coupled {B}ose gas},\ }\href {https://doi.org/10.1038/nphys2905} {\bibfield  {journal} {\bibinfo  {journal} {Nature Physics}\ }\textbf {\bibinfo {volume} {10}},\ \bibinfo {pages} {314} (\bibinfo {year} {2014})}\BibitemShut {NoStop}%
\bibitem [{\citenamefont {Jim\'enez-Garc\'{\i}a}\ \emph {et~al.}(2015)\citenamefont {Jim\'enez-Garc\'{\i}a}, \citenamefont {LeBlanc}, \citenamefont {Williams}, \citenamefont {Beeler}, \citenamefont {Qu}, \citenamefont {Gong}, \citenamefont {Zhang},\ and\ \citenamefont {Spielman}}]{Jimenez-Garcia2015}%
  \BibitemOpen
  \bibfield  {author} {\bibinfo {author} {\bibfnamefont {K.}~\bibnamefont {Jim\'enez-Garc\'{\i}a}}, \bibinfo {author} {\bibfnamefont {L.~J.}\ \bibnamefont {LeBlanc}}, \bibinfo {author} {\bibfnamefont {R.~A.}\ \bibnamefont {Williams}}, \bibinfo {author} {\bibfnamefont {M.~C.}\ \bibnamefont {Beeler}}, \bibinfo {author} {\bibfnamefont {C.}~\bibnamefont {Qu}}, \bibinfo {author} {\bibfnamefont {M.}~\bibnamefont {Gong}}, \bibinfo {author} {\bibfnamefont {C.}~\bibnamefont {Zhang}},\ and\ \bibinfo {author} {\bibfnamefont {I.~B.}\ \bibnamefont {Spielman}},\ }\bibfield  {title} {\bibinfo {title} {Tunable spin-orbit coupling via strong driving in ultracold-atom systems},\ }\href {https://doi.org/10.1103/PhysRevLett.114.125301} {\bibfield  {journal} {\bibinfo  {journal} {Phys. Rev. Lett.}\ }\textbf {\bibinfo {volume} {114}},\ \bibinfo {pages} {125301} (\bibinfo {year} {2015})}\BibitemShut {NoStop}%
\bibitem [{\citenamefont {Grusdt}\ \emph {et~al.}(2017)\citenamefont {Grusdt}, \citenamefont {Li}, \citenamefont {Bloch},\ and\ \citenamefont {Demler}}]{Grusdt2017}%
  \BibitemOpen
  \bibfield  {author} {\bibinfo {author} {\bibfnamefont {F.}~\bibnamefont {Grusdt}}, \bibinfo {author} {\bibfnamefont {T.}~\bibnamefont {Li}}, \bibinfo {author} {\bibfnamefont {I.}~\bibnamefont {Bloch}},\ and\ \bibinfo {author} {\bibfnamefont {E.}~\bibnamefont {Demler}},\ }\bibfield  {title} {\bibinfo {title} {{Tunable spin-orbit coupling for ultracold atoms in two-dimensional optical lattices}},\ }\href {https://doi.org/10.1103/PhysRevA.95.063617} {\bibfield  {journal} {\bibinfo  {journal} {Phys. Rev. A}\ }\textbf {\bibinfo {volume} {95}},\ \bibinfo {pages} {063617} (\bibinfo {year} {2017})}\BibitemShut {NoStop}%
\bibitem [{\citenamefont {Li}\ \emph {et~al.}(2018)\citenamefont {Li}, \citenamefont {Ye}, \citenamefont {Chen}, \citenamefont {Kartashov}, \citenamefont {Torner},\ and\ \citenamefont {Konotop}}]{Li2018}%
  \BibitemOpen
  \bibfield  {author} {\bibinfo {author} {\bibfnamefont {C.}~\bibnamefont {Li}}, \bibinfo {author} {\bibfnamefont {F.}~\bibnamefont {Ye}}, \bibinfo {author} {\bibfnamefont {X.}~\bibnamefont {Chen}}, \bibinfo {author} {\bibfnamefont {Y.~V.}\ \bibnamefont {Kartashov}}, \bibinfo {author} {\bibfnamefont {L.}~\bibnamefont {Torner}},\ and\ \bibinfo {author} {\bibfnamefont {V.~V.}\ \bibnamefont {Konotop}},\ }\bibfield  {title} {\bibinfo {title} {{Topological edge states in Rashba-Dresselhaus spin-orbit-coupled atoms in a Zeeman lattice}},\ }\href {https://doi.org/10.1103/PhysRevA.98.061601} {\bibfield  {journal} {\bibinfo  {journal} {Phys. Rev. A}\ }\textbf {\bibinfo {volume} {98}},\ \bibinfo {pages} {061601} (\bibinfo {year} {2018})}\BibitemShut {NoStop}%
\bibitem [{\citenamefont {Gross}\ and\ \citenamefont {Bloch}(2017)}]{Gross2017}%
  \BibitemOpen
  \bibfield  {author} {\bibinfo {author} {\bibfnamefont {C.}~\bibnamefont {Gross}}\ and\ \bibinfo {author} {\bibfnamefont {I.}~\bibnamefont {Bloch}},\ }\bibfield  {title} {\bibinfo {title} {Quantum simulations with ultracold atoms in optical lattices},\ }\href {https://doi.org/10.1126/science.aal3837} {\bibfield  {journal} {\bibinfo  {journal} {Science}\ }\textbf {\bibinfo {volume} {357}},\ \bibinfo {pages} {995} (\bibinfo {year} {2017})}\BibitemShut {NoStop}%
\bibitem [{\citenamefont {Hirsch}(1985)}]{Hirsch1985}%
  \BibitemOpen
  \bibfield  {author} {\bibinfo {author} {\bibfnamefont {J.~E.}\ \bibnamefont {Hirsch}},\ }\bibfield  {title} {\bibinfo {title} {{Two-dimensional Hubbard model: Numerical simulation study}},\ }\href {https://doi.org/10.1103/PhysRevB.31.4403} {\bibfield  {journal} {\bibinfo  {journal} {Phys. Rev. B}\ }\textbf {\bibinfo {volume} {31}},\ \bibinfo {pages} {4403} (\bibinfo {year} {1985})}\BibitemShut {NoStop}%
\bibitem [{\citenamefont {Paiva}\ \emph {et~al.}(2005)\citenamefont {Paiva}, \citenamefont {Scalettar}, \citenamefont {Zheng}, \citenamefont {Singh},\ and\ \citenamefont {Oitmaa}}]{Paiva2005}%
  \BibitemOpen
  \bibfield  {author} {\bibinfo {author} {\bibfnamefont {T.}~\bibnamefont {Paiva}}, \bibinfo {author} {\bibfnamefont {R.~T.}\ \bibnamefont {Scalettar}}, \bibinfo {author} {\bibfnamefont {W.}~\bibnamefont {Zheng}}, \bibinfo {author} {\bibfnamefont {R.~R.~P.}\ \bibnamefont {Singh}},\ and\ \bibinfo {author} {\bibfnamefont {J.}~\bibnamefont {Oitmaa}},\ }\bibfield  {title} {\bibinfo {title} {Ground-state and finite-temperature signatures of quantum phase transitions in the half-filled {H}ubbard model on a honeycomb lattice},\ }\href {https://doi.org/10.1103/PhysRevB.72.085123} {\bibfield  {journal} {\bibinfo  {journal} {Phys. Rev. B}\ }\textbf {\bibinfo {volume} {72}},\ \bibinfo {pages} {085123} (\bibinfo {year} {2005})}\BibitemShut {NoStop}%
\bibitem [{\citenamefont {Meng}\ \emph {et~al.}(2010)\citenamefont {Meng}, \citenamefont {Lang}, \citenamefont {Wessel}, \citenamefont {Assaad},\ and\ \citenamefont {Muramatsu}}]{Meng2010}%
  \BibitemOpen
  \bibfield  {author} {\bibinfo {author} {\bibfnamefont {Z.~Y.}\ \bibnamefont {Meng}}, \bibinfo {author} {\bibfnamefont {T.~C.}\ \bibnamefont {Lang}}, \bibinfo {author} {\bibfnamefont {S.}~\bibnamefont {Wessel}}, \bibinfo {author} {\bibfnamefont {F.~F.}\ \bibnamefont {Assaad}},\ and\ \bibinfo {author} {\bibfnamefont {A.}~\bibnamefont {Muramatsu}},\ }\bibfield  {title} {\bibinfo {title} {Quantum spin liquid emerging in two-dimensional correlated {D}irac fermions},\ }\href {https://doi.org/10.1038/nature08942} {\bibfield  {journal} {\bibinfo  {journal} {Nature}\ }\textbf {\bibinfo {volume} {464}},\ \bibinfo {pages} {847} (\bibinfo {year} {2010})}\BibitemShut {NoStop}%
\bibitem [{\citenamefont {Sorella}\ \emph {et~al.}(2012)\citenamefont {Sorella}, \citenamefont {Otsuka},\ and\ \citenamefont {Yunoki}}]{Sorella2012}%
  \BibitemOpen
  \bibfield  {author} {\bibinfo {author} {\bibfnamefont {S.}~\bibnamefont {Sorella}}, \bibinfo {author} {\bibfnamefont {Y.}~\bibnamefont {Otsuka}},\ and\ \bibinfo {author} {\bibfnamefont {S.}~\bibnamefont {Yunoki}},\ }\bibfield  {title} {\bibinfo {title} {Absence of a spin liquid phase in the {H}ubbard model on the honeycomb lattice},\ }\href {https://doi.org/10.1038/srep00992} {\bibfield  {journal} {\bibinfo  {journal} {Scientific Reports}\ }\textbf {\bibinfo {volume} {2}},\ \bibinfo {pages} {992} (\bibinfo {year} {2012})}\BibitemShut {NoStop}%
\bibitem [{\citenamefont {Cole}\ \emph {et~al.}(2012)\citenamefont {Cole}, \citenamefont {Zhang}, \citenamefont {Paramekanti},\ and\ \citenamefont {Trivedi}}]{Cole2012}%
  \BibitemOpen
  \bibfield  {author} {\bibinfo {author} {\bibfnamefont {W.~S.}\ \bibnamefont {Cole}}, \bibinfo {author} {\bibfnamefont {S.}~\bibnamefont {Zhang}}, \bibinfo {author} {\bibfnamefont {A.}~\bibnamefont {Paramekanti}},\ and\ \bibinfo {author} {\bibfnamefont {N.}~\bibnamefont {Trivedi}},\ }\bibfield  {title} {\bibinfo {title} {{Bose-Hubbard Models with Synthetic Spin-Orbit Coupling: Mott Insulators, Spin Textures, and Superfluidity}},\ }\href {https://doi.org/10.1103/PhysRevLett.109.085302} {\bibfield  {journal} {\bibinfo  {journal} {Phys. Rev. Lett.}\ }\textbf {\bibinfo {volume} {109}},\ \bibinfo {pages} {085302} (\bibinfo {year} {2012})}\BibitemShut {NoStop}%
\bibitem [{\citenamefont {Min\'a\ifmmode~\check{r}\else \v{r}\fi{}}\ and\ \citenamefont {Gr\'emaud}(2013)}]{Minar2013}%
  \BibitemOpen
  \bibfield  {author} {\bibinfo {author} {\bibfnamefont {J.~c.~v.}\ \bibnamefont {Min\'a\ifmmode~\check{r}\else \v{r}\fi{}}}\ and\ \bibinfo {author} {\bibfnamefont {B.}~\bibnamefont {Gr\'emaud}},\ }\bibfield  {title} {\bibinfo {title} {{From antiferromagnetic ordering to magnetic textures in the two-dimensional Fermi-Hubbard model with synthetic spin-orbit interactions}},\ }\href {https://doi.org/10.1103/PhysRevB.88.235130} {\bibfield  {journal} {\bibinfo  {journal} {Phys. Rev. B}\ }\textbf {\bibinfo {volume} {88}},\ \bibinfo {pages} {235130} (\bibinfo {year} {2013})}\BibitemShut {NoStop}%
\bibitem [{\citenamefont {Wang}\ \emph {et~al.}(2017)\citenamefont {Wang}, \citenamefont {Gong}, \citenamefont {Han}, \citenamefont {Guo},\ and\ \citenamefont {He}}]{Wang2017}%
  \BibitemOpen
  \bibfield  {author} {\bibinfo {author} {\bibfnamefont {C.}~\bibnamefont {Wang}}, \bibinfo {author} {\bibfnamefont {M.}~\bibnamefont {Gong}}, \bibinfo {author} {\bibfnamefont {Y.}~\bibnamefont {Han}}, \bibinfo {author} {\bibfnamefont {G.}~\bibnamefont {Guo}},\ and\ \bibinfo {author} {\bibfnamefont {L.}~\bibnamefont {He}},\ }\bibfield  {title} {\bibinfo {title} {{Exotic spin phases in two-dimensional spin-orbit coupled models: Importance of quantum effects}},\ }\href {https://doi.org/10.1103/PhysRevB.96.115119} {\bibfield  {journal} {\bibinfo  {journal} {Phys. Rev. B}\ }\textbf {\bibinfo {volume} {96}},\ \bibinfo {pages} {115119} (\bibinfo {year} {2017})}\BibitemShut {NoStop}%
\bibitem [{\citenamefont {Kawano}\ and\ \citenamefont {Hotta}(2023)}]{Kawano2023}%
  \BibitemOpen
  \bibfield  {author} {\bibinfo {author} {\bibfnamefont {M.}~\bibnamefont {Kawano}}\ and\ \bibinfo {author} {\bibfnamefont {C.}~\bibnamefont {Hotta}},\ }\bibfield  {title} {\bibinfo {title} {{Phase diagram of the square-lattice Hubbard model with Rashba-type antisymmetric spin-orbit coupling}},\ }\href {https://doi.org/10.1103/PhysRevB.107.045123} {\bibfield  {journal} {\bibinfo  {journal} {Phys. Rev. B}\ }\textbf {\bibinfo {volume} {107}},\ \bibinfo {pages} {045123} (\bibinfo {year} {2023})}\BibitemShut {NoStop}%
\bibitem [{\citenamefont {Kennedy}\ \emph {et~al.}(2022)\citenamefont {Kennedy}, \citenamefont {Sousa-J\'unior}, \citenamefont {Costa},\ and\ \citenamefont {dos Santos}}]{Kennedy2022}%
  \BibitemOpen
  \bibfield  {author} {\bibinfo {author} {\bibfnamefont {W.}~\bibnamefont {Kennedy}}, \bibinfo {author} {\bibfnamefont {S.~a. d.~A.}\ \bibnamefont {Sousa-J\'unior}}, \bibinfo {author} {\bibfnamefont {N.~C.}\ \bibnamefont {Costa}},\ and\ \bibinfo {author} {\bibfnamefont {R.~R.}\ \bibnamefont {dos Santos}},\ }\bibfield  {title} {\bibinfo {title} {{Magnetism and metal-insulator transitions in the Rashba-Hubbard model}},\ }\href {https://doi.org/10.1103/PhysRevB.106.165121} {\bibfield  {journal} {\bibinfo  {journal} {Phys. Rev. B}\ }\textbf {\bibinfo {volume} {106}},\ \bibinfo {pages} {165121} (\bibinfo {year} {2022})}\BibitemShut {NoStop}%
\bibitem [{\citenamefont {Kubo}(2024)}]{Kubo2024}%
  \BibitemOpen
  \bibfield  {author} {\bibinfo {author} {\bibfnamefont {K.}~\bibnamefont {Kubo}},\ }\bibfield  {title} {\bibinfo {title} {{Weyl Semimetallic State in the Rashba–Hubbard Model}},\ }\href {https://doi.org/10.7566/JPSJ.93.024708} {\bibfield  {journal} {\bibinfo  {journal} {Journal of the Physical Society of Japan}\ }\textbf {\bibinfo {volume} {93}},\ \bibinfo {pages} {024708} (\bibinfo {year} {2024})}\BibitemShut {NoStop}%
\bibitem [{\citenamefont {Jain}\ \emph {et~al.}(2024)\citenamefont {Jain}, \citenamefont {Goyal},\ and\ \citenamefont {Singh}}]{jain2024}%
  \BibitemOpen
  \bibfield  {author} {\bibinfo {author} {\bibfnamefont {A.}~\bibnamefont {Jain}}, \bibinfo {author} {\bibfnamefont {G.}~\bibnamefont {Goyal}},\ and\ \bibinfo {author} {\bibfnamefont {D.~K.}\ \bibnamefont {Singh}},\ }\bibfield  {title} {\bibinfo {title} {{Weyl semimetallic state with antiferromagnetic order in the Rashba-Hubbard model}},\ }\href {https://doi.org/10.1103/PhysRevB.110.075134} {\bibfield  {journal} {\bibinfo  {journal} {Phys. Rev. B}\ }\textbf {\bibinfo {volume} {110}},\ \bibinfo {pages} {075134} (\bibinfo {year} {2024})}\BibitemShut {NoStop}%
\bibitem [{\citenamefont {Brosco}\ \emph {et~al.}(2018)\citenamefont {Brosco}, \citenamefont {Guerci},\ and\ \citenamefont {Capone}}]{Brosco2018}%
  \BibitemOpen
  \bibfield  {author} {\bibinfo {author} {\bibfnamefont {V.}~\bibnamefont {Brosco}}, \bibinfo {author} {\bibfnamefont {D.}~\bibnamefont {Guerci}},\ and\ \bibinfo {author} {\bibfnamefont {M.}~\bibnamefont {Capone}},\ }\bibfield  {title} {\bibinfo {title} {{Pauli metallic ground state in Hubbard clusters with Rashba spin-orbit coupling}},\ }\href {https://doi.org/10.1103/PhysRevB.97.125103} {\bibfield  {journal} {\bibinfo  {journal} {Phys. Rev. B}\ }\textbf {\bibinfo {volume} {97}},\ \bibinfo {pages} {125103} (\bibinfo {year} {2018})}\BibitemShut {NoStop}%
\bibitem [{\citenamefont {Brosco}\ and\ \citenamefont {Capone}(2020)}]{Brosco2020}%
  \BibitemOpen
  \bibfield  {author} {\bibinfo {author} {\bibfnamefont {V.}~\bibnamefont {Brosco}}\ and\ \bibinfo {author} {\bibfnamefont {M.}~\bibnamefont {Capone}},\ }\bibfield  {title} {\bibinfo {title} {{Rashba-metal to Mott-insulator transition}},\ }\href {https://doi.org/10.1103/PhysRevB.101.235149} {\bibfield  {journal} {\bibinfo  {journal} {Phys. Rev. B}\ }\textbf {\bibinfo {volume} {101}},\ \bibinfo {pages} {235149} (\bibinfo {year} {2020})}\BibitemShut {NoStop}%
\bibitem [{\citenamefont {Fontenele}\ \emph {et~al.}(2024)\citenamefont {Fontenele}, \citenamefont {dos Anjos Sousa~Júnior}, \citenamefont {Cysne},\ and\ \citenamefont {Costa}}]{Fontenele2024}%
  \BibitemOpen
  \bibfield  {author} {\bibinfo {author} {\bibfnamefont {R.~A.}\ \bibnamefont {Fontenele}}, \bibinfo {author} {\bibfnamefont {S.}~\bibnamefont {dos Anjos Sousa~Júnior}}, \bibinfo {author} {\bibfnamefont {T.~P.}\ \bibnamefont {Cysne}},\ and\ \bibinfo {author} {\bibfnamefont {N.~C.}\ \bibnamefont {Costa}},\ }\bibfield  {title} {\bibinfo {title} {{The impact of Rashba spin-orbit coupling in charge-ordered systems}},\ }\href {https://doi.org/10.1088/1361-648X/ad2cc9} {\bibfield  {journal} {\bibinfo  {journal} {Journal of Physics: Condensed Matter}\ }\textbf {\bibinfo {volume} {36}},\ \bibinfo {pages} {225601} (\bibinfo {year} {2024})}\BibitemShut {NoStop}%
\bibitem [{\citenamefont {Faúndez}\ \emph {et~al.}(2024)\citenamefont {Faúndez}, \citenamefont {Fontenele}, \citenamefont {dos Anjos Sousa-Júnior}, \citenamefont {Assaad},\ and\ \citenamefont {Costa}}]{Faundez2024}%
  \BibitemOpen
  \bibfield  {author} {\bibinfo {author} {\bibfnamefont {J.}~\bibnamefont {Faúndez}}, \bibinfo {author} {\bibfnamefont {R.~A.}\ \bibnamefont {Fontenele}}, \bibinfo {author} {\bibfnamefont {S.}~\bibnamefont {dos Anjos Sousa-Júnior}}, \bibinfo {author} {\bibfnamefont {F.~F.}\ \bibnamefont {Assaad}},\ and\ \bibinfo {author} {\bibfnamefont {N.~C.}\ \bibnamefont {Costa}},\ }\href {https://arxiv.org/abs/2411.07119} {\bibinfo {title} {{The two-dimensional Rashba-Holstein model}}} (\bibinfo {year} {2024}),\ \Eprint {https://arxiv.org/abs/2411.07119} {arXiv:2411.07119 [cond-mat.str-el]} \BibitemShut {NoStop}%
\bibitem [{\citenamefont {Shigeta}\ \emph {et~al.}(2013)\citenamefont {Shigeta}, \citenamefont {Onari},\ and\ \citenamefont {Tanaka}}]{Shigeta2013}%
  \BibitemOpen
  \bibfield  {author} {\bibinfo {author} {\bibfnamefont {K.}~\bibnamefont {Shigeta}}, \bibinfo {author} {\bibfnamefont {S.}~\bibnamefont {Onari}},\ and\ \bibinfo {author} {\bibfnamefont {Y.}~\bibnamefont {Tanaka}},\ }\bibfield  {title} {\bibinfo {title} {Superconducting pairing symmetry on the extended {H}ubbard model in the presence of the {R}ashba-type spin–orbit coupling},\ }\href {https://doi.org/10.7566/JPSJ.82.014702} {\bibfield  {journal} {\bibinfo  {journal} {Journal of the Physical Society of Japan}\ }\textbf {\bibinfo {volume} {82}},\ \bibinfo {pages} {014702} (\bibinfo {year} {2013})}\BibitemShut {NoStop}%
\bibitem [{\citenamefont {Beyer}\ \emph {et~al.}(2023)\citenamefont {Beyer}, \citenamefont {Profe}, \citenamefont {Klebl}, \citenamefont {Schwemmer}, \citenamefont {Kennes}, \citenamefont {Thomale}, \citenamefont {Honerkamp},\ and\ \citenamefont {Rachel}}]{Beyer2023}%
  \BibitemOpen
  \bibfield  {author} {\bibinfo {author} {\bibfnamefont {J.}~\bibnamefont {Beyer}}, \bibinfo {author} {\bibfnamefont {J.~B.}\ \bibnamefont {Profe}}, \bibinfo {author} {\bibfnamefont {L.}~\bibnamefont {Klebl}}, \bibinfo {author} {\bibfnamefont {T.}~\bibnamefont {Schwemmer}}, \bibinfo {author} {\bibfnamefont {D.~M.}\ \bibnamefont {Kennes}}, \bibinfo {author} {\bibfnamefont {R.}~\bibnamefont {Thomale}}, \bibinfo {author} {\bibfnamefont {C.}~\bibnamefont {Honerkamp}},\ and\ \bibinfo {author} {\bibfnamefont {S.}~\bibnamefont {Rachel}},\ }\bibfield  {title} {\bibinfo {title} {{Rashba spin-orbit coupling in the square-lattice Hubbard model: A truncated-unity functional renormalization group study}},\ }\href {https://doi.org/10.1103/PhysRevB.107.125115} {\bibfield  {journal} {\bibinfo  {journal} {Phys. Rev. B}\ }\textbf {\bibinfo {volume} {107}},\ \bibinfo {pages} {125115} (\bibinfo {year} {2023})}\BibitemShut {NoStop}%
\bibitem [{\citenamefont {Tang}\ \emph {et~al.}(2014)\citenamefont {Tang}, \citenamefont {Yang}, \citenamefont {Sun},\ and\ \citenamefont {Lin}}]{Tang2014}%
  \BibitemOpen
  \bibfield  {author} {\bibinfo {author} {\bibfnamefont {H.-K.}\ \bibnamefont {Tang}}, \bibinfo {author} {\bibfnamefont {X.}~\bibnamefont {Yang}}, \bibinfo {author} {\bibfnamefont {J.}~\bibnamefont {Sun}},\ and\ \bibinfo {author} {\bibfnamefont {H.-Q.}\ \bibnamefont {Lin}},\ }\bibfield  {title} {\bibinfo {title} {{Berezinskii-Kosterlitz-Thoules phase transition of spin-orbit coupled Fermi gas in optical lattice}},\ }\href {https://doi.org/10.1209/0295-5075/107/40003} {\bibfield  {journal} {\bibinfo  {journal} {Europhysics Letters}\ }\textbf {\bibinfo {volume} {107}},\ \bibinfo {pages} {40003} (\bibinfo {year} {2014})}\BibitemShut {NoStop}%
\bibitem [{\citenamefont {Rosenberg}\ \emph {et~al.}(2017)\citenamefont {Rosenberg}, \citenamefont {Shi},\ and\ \citenamefont {Zhang}}]{Rosenberg2017}%
  \BibitemOpen
  \bibfield  {author} {\bibinfo {author} {\bibfnamefont {P.}~\bibnamefont {Rosenberg}}, \bibinfo {author} {\bibfnamefont {H.}~\bibnamefont {Shi}},\ and\ \bibinfo {author} {\bibfnamefont {S.}~\bibnamefont {Zhang}},\ }\bibfield  {title} {\bibinfo {title} {{Ultracold Atoms in a Square Lattice with Spin-Orbit Coupling: Charge Order, Superfluidity, and Topological Signatures}},\ }\href {https://doi.org/10.1103/PhysRevLett.119.265301} {\bibfield  {journal} {\bibinfo  {journal} {Phys. Rev. Lett.}\ }\textbf {\bibinfo {volume} {119}},\ \bibinfo {pages} {265301} (\bibinfo {year} {2017})}\BibitemShut {NoStop}%
\bibitem [{\citenamefont {Wan}\ \emph {et~al.}(2022)\citenamefont {Wan}, \citenamefont {Zhang},\ and\ \citenamefont {Yao}}]{Wan2022}%
  \BibitemOpen
  \bibfield  {author} {\bibinfo {author} {\bibfnamefont {Z.-Q.}\ \bibnamefont {Wan}}, \bibinfo {author} {\bibfnamefont {S.-X.}\ \bibnamefont {Zhang}},\ and\ \bibinfo {author} {\bibfnamefont {H.}~\bibnamefont {Yao}},\ }\bibfield  {title} {\bibinfo {title} {{Mitigating the fermion sign problem by automatic differentiation}},\ }\href {https://doi.org/10.1103/PhysRevB.106.L241109} {\bibfield  {journal} {\bibinfo  {journal} {Phys. Rev. B}\ }\textbf {\bibinfo {volume} {106}},\ \bibinfo {pages} {L241109} (\bibinfo {year} {2022})}\BibitemShut {NoStop}%
\bibitem [{\citenamefont {Stewart}(2002)}]{Stewart2002}%
  \BibitemOpen
  \bibfield  {author} {\bibinfo {author} {\bibfnamefont {G.~W.}\ \bibnamefont {Stewart}},\ }\bibfield  {title} {\bibinfo {title} {A {K}rylov--{S}chur algorithm for large eigenproblems},\ }\href {https://doi.org/10.1137/S0895479800371529} {\bibfield  {journal} {\bibinfo  {journal} {SIAM Journal on Matrix Analysis and Applications}\ }\textbf {\bibinfo {volume} {23}},\ \bibinfo {pages} {601} (\bibinfo {year} {2002})}\BibitemShut {NoStop}%
\bibitem [{\citenamefont {Hernandez}\ \emph {et~al.}(2005)\citenamefont {Hernandez}, \citenamefont {Roman},\ and\ \citenamefont {Vidal}}]{Slepc}%
  \BibitemOpen
  \bibfield  {author} {\bibinfo {author} {\bibfnamefont {V.}~\bibnamefont {Hernandez}}, \bibinfo {author} {\bibfnamefont {J.~E.}\ \bibnamefont {Roman}},\ and\ \bibinfo {author} {\bibfnamefont {V.}~\bibnamefont {Vidal}},\ }\bibfield  {title} {\bibinfo {title} {{SLEPc}: A scalable and flexible toolkit for the solution of eigenvalue problems},\ }\href@noop {} {\bibfield  {journal} {\bibinfo  {journal} {{ACM} Trans. Math. Software}\ }\textbf {\bibinfo {volume} {31}},\ \bibinfo {pages} {351} (\bibinfo {year} {2005})}\BibitemShut {NoStop}%
\bibitem [{\citenamefont {Blankenbecler}\ \emph {et~al.}(1981)\citenamefont {Blankenbecler}, \citenamefont {Scalapino},\ and\ \citenamefont {Sugar}}]{Blankenbecler1981}%
  \BibitemOpen
  \bibfield  {author} {\bibinfo {author} {\bibfnamefont {R.}~\bibnamefont {Blankenbecler}}, \bibinfo {author} {\bibfnamefont {D.~J.}\ \bibnamefont {Scalapino}},\ and\ \bibinfo {author} {\bibfnamefont {R.~L.}\ \bibnamefont {Sugar}},\ }\bibfield  {title} {\bibinfo {title} {{Monte Carlo calculations of coupled boson-fermion systems. I}},\ }\href {https://doi.org/10.1103/PhysRevD.24.2278} {\bibfield  {journal} {\bibinfo  {journal} {Phys. Rev. D}\ }\textbf {\bibinfo {volume} {24}},\ \bibinfo {pages} {2278} (\bibinfo {year} {1981})}\BibitemShut {NoStop}%
\bibitem [{\citenamefont {White}\ \emph {et~al.}(1989)\citenamefont {White}, \citenamefont {Scalapino}, \citenamefont {Sugar}, \citenamefont {Loh}, \citenamefont {Gubernatis},\ and\ \citenamefont {Scalettar}}]{White1989}%
  \BibitemOpen
  \bibfield  {author} {\bibinfo {author} {\bibfnamefont {S.~R.}\ \bibnamefont {White}}, \bibinfo {author} {\bibfnamefont {D.~J.}\ \bibnamefont {Scalapino}}, \bibinfo {author} {\bibfnamefont {R.~L.}\ \bibnamefont {Sugar}}, \bibinfo {author} {\bibfnamefont {E.~Y.}\ \bibnamefont {Loh}}, \bibinfo {author} {\bibfnamefont {J.~E.}\ \bibnamefont {Gubernatis}},\ and\ \bibinfo {author} {\bibfnamefont {R.~T.}\ \bibnamefont {Scalettar}},\ }\bibfield  {title} {\bibinfo {title} {{Numerical study of the two-dimensional Hubbard model}},\ }\href {https://doi.org/10.1103/PhysRevB.40.506} {\bibfield  {journal} {\bibinfo  {journal} {Phys. Rev. B}\ }\textbf {\bibinfo {volume} {40}},\ \bibinfo {pages} {506} (\bibinfo {year} {1989})}\BibitemShut {NoStop}%
\bibitem [{\citenamefont {Hirsch}(1983)}]{Hirsch1983}%
  \BibitemOpen
  \bibfield  {author} {\bibinfo {author} {\bibfnamefont {J.~E.}\ \bibnamefont {Hirsch}},\ }\bibfield  {title} {\bibinfo {title} {{Discrete Hubbard-Stratonovich transformation for fermion lattice models}},\ }\href {https://doi.org/10.1103/PhysRevB.28.4059} {\bibfield  {journal} {\bibinfo  {journal} {Phys. Rev. B}\ }\textbf {\bibinfo {volume} {28}},\ \bibinfo {pages} {4059} (\bibinfo {year} {1983})}\BibitemShut {NoStop}%
\bibitem [{\citenamefont {dos Santos}(2003)}]{rrds2003}%
  \BibitemOpen
  \bibfield  {author} {\bibinfo {author} {\bibfnamefont {R.~R.}\ \bibnamefont {dos Santos}},\ }\bibfield  {title} {\bibinfo {title} {{Introduction to quantum Monte Carlo simulations for fermionic systems}},\ }\href {https://doi.org/10.1590/S0103-97332003000100003} {\bibfield  {journal} {\bibinfo  {journal} {Braz. J. Phys}\ }\textbf {\bibinfo {volume} {33}},\ \bibinfo {pages} {63} (\bibinfo {year} {2003})}\BibitemShut {NoStop}%
\bibitem [{\citenamefont {White}\ and\ \citenamefont {Wilkins}(1988)}]{White1988}%
  \BibitemOpen
  \bibfield  {author} {\bibinfo {author} {\bibfnamefont {S.~R.}\ \bibnamefont {White}}\ and\ \bibinfo {author} {\bibfnamefont {J.~W.}\ \bibnamefont {Wilkins}},\ }\bibfield  {title} {\bibinfo {title} {{Fermion simulations in systems with negative weights}},\ }\href {https://doi.org/10.1103/PhysRevB.37.5024} {\bibfield  {journal} {\bibinfo  {journal} {Phys. Rev. B}\ }\textbf {\bibinfo {volume} {37}},\ \bibinfo {pages} {5024} (\bibinfo {year} {1988})}\BibitemShut {NoStop}%
\bibitem [{\citenamefont {Loh}\ \emph {et~al.}(1990)\citenamefont {Loh}, \citenamefont {Gubernatis}, \citenamefont {Scalettar}, \citenamefont {White}, \citenamefont {Scalapino},\ and\ \citenamefont {Sugar}}]{Loh1990}%
  \BibitemOpen
  \bibfield  {author} {\bibinfo {author} {\bibfnamefont {E.~Y.}\ \bibnamefont {Loh}}, \bibinfo {author} {\bibfnamefont {J.~E.}\ \bibnamefont {Gubernatis}}, \bibinfo {author} {\bibfnamefont {R.~T.}\ \bibnamefont {Scalettar}}, \bibinfo {author} {\bibfnamefont {S.~R.}\ \bibnamefont {White}}, \bibinfo {author} {\bibfnamefont {D.~J.}\ \bibnamefont {Scalapino}},\ and\ \bibinfo {author} {\bibfnamefont {R.~L.}\ \bibnamefont {Sugar}},\ }\bibfield  {title} {\bibinfo {title} {{Sign problem in the numerical simulation of many-electron systems}},\ }\href {https://doi.org/10.1103/PhysRevB.41.9301} {\bibfield  {journal} {\bibinfo  {journal} {Phys. Rev. B}\ }\textbf {\bibinfo {volume} {41}},\ \bibinfo {pages} {9301} (\bibinfo {year} {1990})}\BibitemShut {NoStop}%
\bibitem [{\citenamefont {Mondaini}\ \emph {et~al.}(2022)\citenamefont {Mondaini}, \citenamefont {Tarat},\ and\ \citenamefont {Scalettar}}]{Mondaini2022}%
  \BibitemOpen
  \bibfield  {author} {\bibinfo {author} {\bibfnamefont {R.}~\bibnamefont {Mondaini}}, \bibinfo {author} {\bibfnamefont {S.}~\bibnamefont {Tarat}},\ and\ \bibinfo {author} {\bibfnamefont {R.~T.}\ \bibnamefont {Scalettar}},\ }\bibfield  {title} {\bibinfo {title} {{Quantum critical points and the sign problem}},\ }\href {https://doi.org/10.1126/science.abg9299} {\bibfield  {journal} {\bibinfo  {journal} {Science}\ }\textbf {\bibinfo {volume} {375}},\ \bibinfo {pages} {418} (\bibinfo {year} {2022})}\BibitemShut {NoStop}%
\bibitem [{\citenamefont {Mondaini}\ \emph {et~al.}(2023)\citenamefont {Mondaini}, \citenamefont {Tarat},\ and\ \citenamefont {Scalettar}}]{Mondaini2023}%
  \BibitemOpen
  \bibfield  {author} {\bibinfo {author} {\bibfnamefont {R.}~\bibnamefont {Mondaini}}, \bibinfo {author} {\bibfnamefont {S.}~\bibnamefont {Tarat}},\ and\ \bibinfo {author} {\bibfnamefont {R.~T.}\ \bibnamefont {Scalettar}},\ }\bibfield  {title} {\bibinfo {title} {{Universality and critical exponents of the fermion sign problem}},\ }\href {https://doi.org/10.1103/PhysRevB.107.245144} {\bibfield  {journal} {\bibinfo  {journal} {Phys. Rev. B}\ }\textbf {\bibinfo {volume} {107}},\ \bibinfo {pages} {245144} (\bibinfo {year} {2023})}\BibitemShut {NoStop}%
\bibitem [{\citenamefont {Parisen~Toldin}\ \emph {et~al.}(2015)\citenamefont {Parisen~Toldin}, \citenamefont {Hohenadler}, \citenamefont {Assaad},\ and\ \citenamefont {Herbut}}]{Parisen2015}%
  \BibitemOpen
  \bibfield  {author} {\bibinfo {author} {\bibfnamefont {F.}~\bibnamefont {Parisen~Toldin}}, \bibinfo {author} {\bibfnamefont {M.}~\bibnamefont {Hohenadler}}, \bibinfo {author} {\bibfnamefont {F.~F.}\ \bibnamefont {Assaad}},\ and\ \bibinfo {author} {\bibfnamefont {I.~F.}\ \bibnamefont {Herbut}},\ }\bibfield  {title} {\bibinfo {title} {Fermionic quantum criticality in honeycomb and $\ensuremath{\pi}$-flux {H}ubbard models: Finite-size scaling of renormalization-group-invariant observables from quantum {M}onte {C}arlo},\ }\href {https://doi.org/10.1103/PhysRevB.91.165108} {\bibfield  {journal} {\bibinfo  {journal} {Phys. Rev. B}\ }\textbf {\bibinfo {volume} {91}},\ \bibinfo {pages} {165108} (\bibinfo {year} {2015})}\BibitemShut {NoStop}%
\bibitem [{\citenamefont {Otsuka}\ \emph {et~al.}(2016)\citenamefont {Otsuka}, \citenamefont {Yunoki},\ and\ \citenamefont {Sorella}}]{Otsuka2016}%
  \BibitemOpen
  \bibfield  {author} {\bibinfo {author} {\bibfnamefont {Y.}~\bibnamefont {Otsuka}}, \bibinfo {author} {\bibfnamefont {S.}~\bibnamefont {Yunoki}},\ and\ \bibinfo {author} {\bibfnamefont {S.}~\bibnamefont {Sorella}},\ }\bibfield  {title} {\bibinfo {title} {{Universal Quantum Criticality in the Metal-Insulator Transition of Two-Dimensional Interacting {D}irac Electrons}},\ }\href {https://doi.org/10.1103/PhysRevX.6.011029} {\bibfield  {journal} {\bibinfo  {journal} {Phys. Rev. X}\ }\textbf {\bibinfo {volume} {6}},\ \bibinfo {pages} {011029} (\bibinfo {year} {2016})}\BibitemShut {NoStop}%
\bibitem [{\citenamefont {MacDonald}\ \emph {et~al.}(1988)\citenamefont {MacDonald}, \citenamefont {Girvin},\ and\ \citenamefont {Yoshioka}}]{MacDonald1988}%
  \BibitemOpen
  \bibfield  {author} {\bibinfo {author} {\bibfnamefont {A.~H.}\ \bibnamefont {MacDonald}}, \bibinfo {author} {\bibfnamefont {S.~M.}\ \bibnamefont {Girvin}},\ and\ \bibinfo {author} {\bibfnamefont {D.}~\bibnamefont {Yoshioka}},\ }\bibfield  {title} {\bibinfo {title} {$\frac{t}{U}$ expansion for the {H}ubbard model},\ }\href {https://doi.org/10.1103/PhysRevB.37.9753} {\bibfield  {journal} {\bibinfo  {journal} {Phys. Rev. B}\ }\textbf {\bibinfo {volume} {37}},\ \bibinfo {pages} {9753} (\bibinfo {year} {1988})}\BibitemShut {NoStop}%
\bibitem [{\citenamefont {Fisher}(1971)}]{Fisher71}%
  \BibitemOpen
  \bibfield  {author} {\bibinfo {author} {\bibfnamefont {M.~E.}\ \bibnamefont {Fisher}},\ }\bibfield  {title} {\bibinfo {title} {Critical {P}henomena},\ }in\ \href@noop {} {\emph {\bibinfo {booktitle} {{P}roceedings of the {E}nrico {F}ermi {I}nternational {S}chool of {P}hysics}}},\ Vol.~\bibinfo {volume} {51},\ \bibinfo {editor} {edited by\ \bibinfo {editor} {\bibfnamefont {M.~S.}\ \bibnamefont {Green}}}\ (\bibinfo  {publisher} {Academic Press, New York},\ \bibinfo {year} {1971})\BibitemShut {NoStop}%
\bibitem [{\citenamefont {dos Santos}\ and\ \citenamefont {Sneddon}(1981)}]{dosSantos81a}%
  \BibitemOpen
  \bibfield  {author} {\bibinfo {author} {\bibfnamefont {R.~R.}\ \bibnamefont {dos Santos}}\ and\ \bibinfo {author} {\bibfnamefont {L.}~\bibnamefont {Sneddon}},\ }\bibfield  {title} {\bibinfo {title} {{Finite-size rescaling transformations}},\ }\href {https://doi.org/10.1103/PhysRevB.23.3541} {\bibfield  {journal} {\bibinfo  {journal} {Phys. Rev. B}\ }\textbf {\bibinfo {volume} {23}},\ \bibinfo {pages} {3541} (\bibinfo {year} {1981})}\BibitemShut {NoStop}%
\bibitem [{\citenamefont {Barber}(1983)}]{Barber83}%
  \BibitemOpen
  \bibfield  {author} {\bibinfo {author} {\bibfnamefont {M.~N.}\ \bibnamefont {Barber}},\ }\bibfield  {title} {\bibinfo {title} {{Finite-size Scaling}},\ }in\ \href@noop {} {\emph {\bibinfo {booktitle} {Phase Transitions and Critical Phenomena}}},\ Vol.~\bibinfo {volume} {8},\ \bibinfo {editor} {edited by\ \bibinfo {editor} {\bibfnamefont {C.}~\bibnamefont {Domb}}\ and\ \bibinfo {editor} {\bibfnamefont {J.~L.}\ \bibnamefont {Lebowitz}}}\ (\bibinfo  {publisher} {Academic Press},\ \bibinfo {address} {New York},\ \bibinfo {year} {1983})\ p.\ \bibinfo {pages} {145}\BibitemShut {NoStop}%
\bibitem [{\citenamefont {Kaul}(2015)}]{Kaul2015}%
  \BibitemOpen
  \bibfield  {author} {\bibinfo {author} {\bibfnamefont {R.~K.}\ \bibnamefont {Kaul}},\ }\bibfield  {title} {\bibinfo {title} {{Spin Nematics, Valence-Bond Solids, and Spin Liquids in {$\mathrm{SO}(N)$} Quantum Spin Models on the Triangular Lattice}},\ }\href {https://doi.org/10.1103/PhysRevLett.115.157202} {\bibfield  {journal} {\bibinfo  {journal} {Phys. Rev. Lett.}\ }\textbf {\bibinfo {volume} {115}},\ \bibinfo {pages} {157202} (\bibinfo {year} {2015})}\BibitemShut {NoStop}%
\bibitem [{\citenamefont {Niu}\ \emph {et~al.}(1985)\citenamefont {Niu}, \citenamefont {Thouless},\ and\ \citenamefont {Wu}}]{Niu1985}%
  \BibitemOpen
  \bibfield  {author} {\bibinfo {author} {\bibfnamefont {Q.}~\bibnamefont {Niu}}, \bibinfo {author} {\bibfnamefont {D.~J.}\ \bibnamefont {Thouless}},\ and\ \bibinfo {author} {\bibfnamefont {Y.-S.}\ \bibnamefont {Wu}},\ }\bibfield  {title} {\bibinfo {title} {Quantized {H}all conductance as a topological invariant},\ }\href {https://doi.org/10.1103/PhysRevB.31.3372} {\bibfield  {journal} {\bibinfo  {journal} {Phys. Rev. B}\ }\textbf {\bibinfo {volume} {31}},\ \bibinfo {pages} {3372} (\bibinfo {year} {1985})}\BibitemShut {NoStop}%
\bibitem [{\citenamefont {Poilblanc}(1991)}]{Poilblanc1991}%
  \BibitemOpen
  \bibfield  {author} {\bibinfo {author} {\bibfnamefont {D.}~\bibnamefont {Poilblanc}},\ }\bibfield  {title} {\bibinfo {title} {{Twisted boundary conditions in cluster calculations of the optical conductivity in two-dimensional lattice models}},\ }\href {https://doi.org/10.1103/PhysRevB.44.9562} {\bibfield  {journal} {\bibinfo  {journal} {Phys. Rev. B}\ }\textbf {\bibinfo {volume} {44}},\ \bibinfo {pages} {9562} (\bibinfo {year} {1991})}\BibitemShut {NoStop}%
\bibitem [{\citenamefont {Gros}(1992)}]{Gros1992}%
  \BibitemOpen
  \bibfield  {author} {\bibinfo {author} {\bibfnamefont {C.}~\bibnamefont {Gros}},\ }\bibfield  {title} {\bibinfo {title} {{The boundary condition integration technique: results for the Hubbard model in 1D and 2D}},\ }\href {https://doi.org/10.1007/BF01323728} {\bibfield  {journal} {\bibinfo  {journal} {Zeitschrift f{\"u}r Physik B Condensed Matter}\ }\textbf {\bibinfo {volume} {86}},\ \bibinfo {pages} {359} (\bibinfo {year} {1992})}\BibitemShut {NoStop}%
\bibitem [{\citenamefont {Zhang}\ \emph {et~al.}(2015)\citenamefont {Zhang}, \citenamefont {Wu}, \citenamefont {Li}, \citenamefont {Wen}, \citenamefont {Sun},\ and\ \citenamefont {Ji}}]{Zhang2015}%
  \BibitemOpen
  \bibfield  {author} {\bibinfo {author} {\bibfnamefont {X.}~\bibnamefont {Zhang}}, \bibinfo {author} {\bibfnamefont {W.}~\bibnamefont {Wu}}, \bibinfo {author} {\bibfnamefont {G.}~\bibnamefont {Li}}, \bibinfo {author} {\bibfnamefont {L.}~\bibnamefont {Wen}}, \bibinfo {author} {\bibfnamefont {Q.}~\bibnamefont {Sun}},\ and\ \bibinfo {author} {\bibfnamefont {A.-C.}\ \bibnamefont {Ji}},\ }\bibfield  {title} {\bibinfo {title} {Phase diagram of interacting fermi gas in spin–orbit coupled square lattices},\ }\href {https://doi.org/10.1088/1367-2630/17/7/073036} {\bibfield  {journal} {\bibinfo  {journal} {New Journal of Physics}\ }\textbf {\bibinfo {volume} {17}},\ \bibinfo {pages} {073036} (\bibinfo {year} {2015})}\BibitemShut {NoStop}%
\bibitem [{\citenamefont {\ifmmode~\check{S}\else \v{S}\fi{}untajs}\ \emph {et~al.}(2020)\citenamefont {\ifmmode~\check{S}\else \v{S}\fi{}untajs}, \citenamefont {Bon\ifmmode~\check{c}\else \v{c}\fi{}a}, \citenamefont {Prosen},\ and\ \citenamefont {Vidmar}}]{Suntajs2020}%
  \BibitemOpen
  \bibfield  {author} {\bibinfo {author} {\bibfnamefont {J.}~\bibnamefont {\ifmmode~\check{S}\else \v{S}\fi{}untajs}}, \bibinfo {author} {\bibfnamefont {J.}~\bibnamefont {Bon\ifmmode~\check{c}\else \v{c}\fi{}a}}, \bibinfo {author} {\bibfnamefont {T.~c.~v.}\ \bibnamefont {Prosen}},\ and\ \bibinfo {author} {\bibfnamefont {L.}~\bibnamefont {Vidmar}},\ }\bibfield  {title} {\bibinfo {title} {{Ergodicity breaking transition in finite disordered spin chains}},\ }\href {https://doi.org/10.1103/PhysRevB.102.064207} {\bibfield  {journal} {\bibinfo  {journal} {Phys. Rev. B}\ }\textbf {\bibinfo {volume} {102}},\ \bibinfo {pages} {064207} (\bibinfo {year} {2020})}\BibitemShut {NoStop}%
\bibitem [{\citenamefont {Assaad}\ and\ \citenamefont {Herbut}(2013)}]{Assaad2013}%
  \BibitemOpen
  \bibfield  {author} {\bibinfo {author} {\bibfnamefont {F.~F.}\ \bibnamefont {Assaad}}\ and\ \bibinfo {author} {\bibfnamefont {I.~F.}\ \bibnamefont {Herbut}},\ }\bibfield  {title} {\bibinfo {title} {{Pinning the Order: The Nature of Quantum Criticality in the {H}ubbard Model on Honeycomb Lattice}},\ }\href {https://doi.org/10.1103/PhysRevX.3.031010} {\bibfield  {journal} {\bibinfo  {journal} {Phys. Rev. X}\ }\textbf {\bibinfo {volume} {3}},\ \bibinfo {pages} {031010} (\bibinfo {year} {2013})}\BibitemShut {NoStop}%
\bibitem [{\citenamefont {Tang}\ \emph {et~al.}(2018)\citenamefont {Tang}, \citenamefont {Leaw}, \citenamefont {Rodrigues}, \citenamefont {Herbut}, \citenamefont {Sengupta}, \citenamefont {Assaad},\ and\ \citenamefont {Adam}}]{Tang2018}%
  \BibitemOpen
  \bibfield  {author} {\bibinfo {author} {\bibfnamefont {H.-K.}\ \bibnamefont {Tang}}, \bibinfo {author} {\bibfnamefont {J.}~\bibnamefont {Leaw}}, \bibinfo {author} {\bibfnamefont {J.}~\bibnamefont {Rodrigues}}, \bibinfo {author} {\bibfnamefont {I.}~\bibnamefont {Herbut}}, \bibinfo {author} {\bibfnamefont {P.}~\bibnamefont {Sengupta}}, \bibinfo {author} {\bibfnamefont {F.}~\bibnamefont {Assaad}},\ and\ \bibinfo {author} {\bibfnamefont {S.}~\bibnamefont {Adam}},\ }\bibfield  {title} {\bibinfo {title} {{The role of electron-electron interactions in two-dimensional Dirac fermions}},\ }\href@noop {} {\bibfield  {journal} {\bibinfo  {journal} {Science}\ }\textbf {\bibinfo {volume} {361}},\ \bibinfo {pages} {570} (\bibinfo {year} {2018})}\BibitemShut {NoStop}%
\bibitem [{\citenamefont {Mou}\ \emph {et~al.}(2022)\citenamefont {Mou}, \citenamefont {Mondaini},\ and\ \citenamefont {Scalettar}}]{Mou2022}%
  \BibitemOpen
  \bibfield  {author} {\bibinfo {author} {\bibfnamefont {Y.}~\bibnamefont {Mou}}, \bibinfo {author} {\bibfnamefont {R.}~\bibnamefont {Mondaini}},\ and\ \bibinfo {author} {\bibfnamefont {R.~T.}\ \bibnamefont {Scalettar}},\ }\bibfield  {title} {\bibinfo {title} {Bilayer {H}ubbard model: Analysis based on the fermionic sign problem},\ }\href {https://doi.org/10.1103/PhysRevB.106.125116} {\bibfield  {journal} {\bibinfo  {journal} {Phys. Rev. B}\ }\textbf {\bibinfo {volume} {106}},\ \bibinfo {pages} {125116} (\bibinfo {year} {2022})}\BibitemShut {NoStop}%
\bibitem [{\citenamefont {S\'en\'echal}\ \emph {et~al.}(2000)\citenamefont {S\'en\'echal}, \citenamefont {Perez},\ and\ \citenamefont {Pioro-Ladri\`ere}}]{Senechal2000}%
  \BibitemOpen
  \bibfield  {author} {\bibinfo {author} {\bibfnamefont {D.}~\bibnamefont {S\'en\'echal}}, \bibinfo {author} {\bibfnamefont {D.}~\bibnamefont {Perez}},\ and\ \bibinfo {author} {\bibfnamefont {M.}~\bibnamefont {Pioro-Ladri\`ere}},\ }\bibfield  {title} {\bibinfo {title} {{Spectral Weight of the Hubbard Model through Cluster Perturbation Theory}},\ }\href {https://doi.org/10.1103/PhysRevLett.84.522} {\bibfield  {journal} {\bibinfo  {journal} {Phys. Rev. Lett.}\ }\textbf {\bibinfo {volume} {84}},\ \bibinfo {pages} {522} (\bibinfo {year} {2000})}\BibitemShut {NoStop}%
\bibitem [{\citenamefont {S\'en\'echal}\ \emph {et~al.}(2002)\citenamefont {S\'en\'echal}, \citenamefont {Perez},\ and\ \citenamefont {Plouffe}}]{Senechal2002}%
  \BibitemOpen
  \bibfield  {author} {\bibinfo {author} {\bibfnamefont {D.}~\bibnamefont {S\'en\'echal}}, \bibinfo {author} {\bibfnamefont {D.}~\bibnamefont {Perez}},\ and\ \bibinfo {author} {\bibfnamefont {D.}~\bibnamefont {Plouffe}},\ }\bibfield  {title} {\bibinfo {title} {Cluster perturbation theory for hubbard models},\ }\href {https://doi.org/10.1103/PhysRevB.66.075129} {\bibfield  {journal} {\bibinfo  {journal} {Phys. Rev. B}\ }\textbf {\bibinfo {volume} {66}},\ \bibinfo {pages} {075129} (\bibinfo {year} {2002})}\BibitemShut {NoStop}%
\bibitem [{\citenamefont {Fukui}\ \emph {et~al.}(2005)\citenamefont {Fukui}, \citenamefont {Hatsugai},\ and\ \citenamefont {Suzuki}}]{Fukui2005}%
  \BibitemOpen
  \bibfield  {author} {\bibinfo {author} {\bibfnamefont {T.}~\bibnamefont {Fukui}}, \bibinfo {author} {\bibfnamefont {Y.}~\bibnamefont {Hatsugai}},\ and\ \bibinfo {author} {\bibfnamefont {H.}~\bibnamefont {Suzuki}},\ }\bibfield  {title} {\bibinfo {title} {Chern numbers in discretized brillouin zone: Efficient method of computing (spin) hall conductances},\ }\href {https://doi.org/10.1143/JPSJ.74.1674} {\bibfield  {journal} {\bibinfo  {journal} {Journal of the Physical Society of Japan}\ }\textbf {\bibinfo {volume} {74}},\ \bibinfo {pages} {1674} (\bibinfo {year} {2005})}\BibitemShut {NoStop}%
\bibitem [{\citenamefont {Varney}\ \emph {et~al.}(2011)\citenamefont {Varney}, \citenamefont {Sun}, \citenamefont {Rigol},\ and\ \citenamefont {Galitski}}]{Varney2011}%
  \BibitemOpen
  \bibfield  {author} {\bibinfo {author} {\bibfnamefont {C.~N.}\ \bibnamefont {Varney}}, \bibinfo {author} {\bibfnamefont {K.}~\bibnamefont {Sun}}, \bibinfo {author} {\bibfnamefont {M.}~\bibnamefont {Rigol}},\ and\ \bibinfo {author} {\bibfnamefont {V.}~\bibnamefont {Galitski}},\ }\bibfield  {title} {\bibinfo {title} {Topological phase transitions for interacting finite systems},\ }\href {https://doi.org/10.1103/PhysRevB.84.241105} {\bibfield  {journal} {\bibinfo  {journal} {Phys. Rev. B}\ }\textbf {\bibinfo {volume} {84}},\ \bibinfo {pages} {241105} (\bibinfo {year} {2011})}\BibitemShut {NoStop}%
\end{thebibliography}%
%%%%%%%%%%%%%%%%%%%%%%%%%%%%%%%%%%%%%%%%%%%%%%%%%%%%%%%%%%%%%%%%%%%%%%%%%%%%%%%%%%%%%

\end{document}